\documentclass[%
 article,
 amsmath,amssymb,
 aps,
]{revtex4-1}

\usepackage{dcolumn}
\usepackage{bm}
\usepackage{color} 
\usepackage{soul} 
\usepackage{graphicx}

\begin{document}

\preprint{APS/123-QED}

\title{Non-perturbative guiding center and stochastic gyrocenter transformations:\\ gyro-phase is the \emph{Kaluza-Klein} $5^{th}$ dimension also for reconciling\\  General Relativity with Quantum Mechanics}

\author{Claudio Di Troia}

 \altaffiliation[At ]{ENEA, dipartimento FSN, via E. Fermi 45 -00044- Frascati (Rome), Italy}
 \email{claudio.ditroia@enea.it}


\begin{abstract}
The non perturbative guiding center transformation is extended to the relativistic regime and takes into account electromagnetic fluctuations.  The main  solutions are obtained in covariant form: the gyrating particle and the guiding particle solutions, both in gyro-kinetic as in MHD orderings.  Moreover,  the presence of a gravitational field is also considered. The way to introduce the gravitational field is original and based on the Einstein conjecture on the feasibility to extend the general relativity theory to include electromagnetism by geometry, if applied to the extended phase space.  In gyro-kinetic theory, some interesting novelties  appear in a natural way, such as the exactness of the conservation of a magnetic moment, or the fact that the gyro-phase is treated as the non observable fifth dimension of the \emph{Kaluza-Klein} model. Electrodynamic becomes non local, without the inconsistency of self-energy. Finally, the gyrocenter transformation is considered in the presence of stochastic e.m. fluctuations for explaining quantum behaviors via Nelson's approach. The gyrocenter law of motion is the \emph{Schr\"odinger} equation.
\end{abstract}

\maketitle

\section{Introduction}
In plasma physics, the gyrokinetic codes are heavily used because they offer the possibility to understand plasma mechanisms from first principles. The collective dynamic is the effect of the self-consistent interaction of single particles with electromagnetic fields. The particle interaction with electromagnetic (e.m.) fields is described by the \emph{Lorentz's force law}, whilst the e.m. fields are described  by Maxwell's equations. The difficulty is in the nonlinearity of the problem, because the same e.m. fields that influence the motion of the single particle are sustained by the four-current charge density made by the same particles.\\
The lagrangian for describing electrodynamics is the sum of the single particle lagrangian, $\ell(t,x,v)$, times the distribution function of particles, $f(t,x,v)$, with  the e.m. lagrangian. The action is often expressed as \cite{turchetti14}:
\begin{equation}
\label{splasma}
S_\mathrm{plasma}=\int  d\,t d\,x d\,v f(t,x,v) \ell(t,x,v)- \int  \frac{F_{\alpha \beta}F^{\alpha \beta}}{4}  d\,t d\,x,
\end{equation}
 where $F_{\alpha \beta}$ is the e.m. tensor.
 This problem is so difficult that some approximations are often considered: the motion of the particles is approximated, \emph{e.g.} in laboratory plasmas the relativistic effects are neglected and/or the non-uniformity of the magnetic field is ignored.  In the present work we use a \emph{non}-perturbative approach for describing the particle relativistic motion in a self-consistent e.m. field. Moreover, mainly for astrophysical and cosmic plasmas, the present description is extended to a \emph{general relativistic} formulation when the presence of a gravitational field is not negligible.  It is worth noticing that  the solution of an exact \emph{Vlasov-Maxwell-Einstein} system gives the most complete description of what concerns the \emph{classical} field theory approach for studying plasmas.\\
The work is divided in four parts. In the first part the single particle lagrangian and its Euler-Lagrangian (EL) equations of motion, i.e. the \emph{Lorentz' force law}, is studied. In the second part the non-perturbative guiding center description is described, which differs a lot from the standard perturbative approach \cite{carybriz}, for obtaining the solutions of the Lorentz' force law.  \\
In the third part it is proposed a method for describing electrodynamics within the general relativity, also for solving the problem of the self-energy. Finally, in the fourth part of this work, electromagnetic fluctuations are considered for obtaining the gyrocenter transformation. The e.m. fluctuations are, firstly, considered as stochastic and the present derivation of the gyrocenter transformation is very different from \cite{brizard}. Once fluctuations are considered it will be possible to include quantum effects through the Nelson's approach (if applied to the guiding center instead of the particle). The introduction of the stochastic calculus, even if necessary, doesn't mean that there are some changes on the physical laws. The Lorentz's force law could remain valid also at a micro-scale. The result is very ambitious because from totally \emph{classical} assumptions it will be possible to propose an explanation of gravitation, electromagnetism and, at least, some aspects of quantum mechanics within the same framework of gyrokinetics applied to general relativity. \\
We can begin by noticing  that there is an asymmetry  in the action (\ref{splasma}) between the particle part and the field part. The e.m. action is obtained  by integrating the lagrangian density over a definite  portion of space-time. This is because \emph{Faraday} defined a field as an object that depends only on space-time variables, e.g. the magnetic field is $B=B(t,x)$. Differently, in the particle action, the motion of charges is described on the whole phase-space during time evolution. The integration is done over the extended phase-space (the phase space plus time). In principle, for restoring the symmetry between the two lagrangians, matter plus fields, it should be simple to think at an action written as
\begin{equation}
\label{somenew}
S_\mathrm{plasma}=\int  d\,t d\,x d\,v \mathcal{L}_\mathrm{plasma},
\end{equation} 
where $ \mathcal{L}_\mathrm{plasma}=f(t,x,v) \ell(t,x,v)+\mathrm{"somethingnew"}$ and the property that 
\begin{equation}
\int \mathrm{"somethingnew"}   d\,v=-\frac{F_{\alpha \beta} F^{\alpha \beta}}{4}.
\end{equation}
Introducing the \emph{phase space lagrangian}, which is a lagrangian density over the extended phase-space, $\mathcal{L}_\mathrm{plasma}=\mathcal{L}_\mathrm{plasma}(t,x,v, \dot{x},\dot{v})$, it could be possible to extend to the whole extended phase-space a $(6+1)$ dimensional  field theory machinery for studying plasmas.\\
In the  theory of gravitation, a similar symmetry  between fields and masses is obtained because the required integration of the lagrangian density is only on a definite portion of space-time, thus the velocity doesn't effectively matter. The gravitational force doesn't depend on the velocity of masses even if gravitation determines the motion of masses, thus, also their velocities. In the lagrangian of a neutral massive body, there is not an interaction term like $A\cdot v$, depending on the velocity of the body. In general relativity theory, it is possible to think at a consistency between the gravitational field and the motion of masses. Indeed, what is said is that the space-time coincides with the gravitational field in the general relativity theory thanks to the Einstein's equation.  The mass trajectory, the curve in space occupied by the mass during time evolution, is a  geodesic on the space-time manifold curved by the presence of masses: the mass can only follow its trajectory consistently with the underlying gravitational field. Is it possible to think at the charge trajectory in a similar fashion? Is it possible to say that the charge trajectory, the curve in phase-space occupied by the charge during time evolution, is the geodesic on the extended phase-space curved by the presence of charges? If yes then it should be possible to obtain an Einstein's equation also for electromagnetism.  \\
The reason for reviewing some topics of the general relativity theory is that in the third part of the present work we will encounter an Hilbert-Einstein (HE) action, as done in the variational approach for deriving Einstein's equation in general relativity but, this is done by considering a metric on the whole extended phase-space. It is proposed to substitute the term "somethingnew" in (\ref{somenew})  with a HE term when velocities are considered as dynamical variables. In this way, we are able to obtain the self-consistent solution of the problem of electrodynamics concerning plasmas in a general e.m. field. Moreover, having used an HE action we will discover that our solutions are also valid in the presence of a gravitational field. If the correctness of such approach will be confirmed the result is very important because it could be said, from now on,  that  the gravitational field coincides with the extended phase-space and not only with the space-time. The important difference with the standard approach is that from giving a geometry to the extended phase-space it is possible to obtain both gravitation with electromagnetism. \\
 Although an Einstein's equation on the extended phase-space should be, somehow, analyzed, it will not been studied here. However, it will be analyzed  what happens if the (non perturbative) guiding center description of motion is adopted.  In such case, a similar mechanism to the one proposed by Kaluza and Klein (KK) a century ago \cite{kaluza, klein} is found. The advantage of using the present description is that, now, there is no need of looking for a compactification scheme as required in the original KK mechanism. Indeed, the extra-dimension that appears in the guiding center transformation is a physical and, in principle, measurable variable being the gyro-phase, the angle obtained when the velocity space is described in a sort of cylindrical transformation of velocities coordinates. Regardless of the equations that are really similar to the one seen in the KK mechanism, the new claim is in the interpretation of the extra dimension as a coordinate coming from the phase-space. Until now, all the compactification mechanisms have been shown to give problems, like the inconsistency of the scale of masses with observations. Instead, without a compactification at the Planck scale length and giving a physical meaning to the extra-coordinate, it seems that the KK mechanism can finally be accepted as a realistic explanation of the presence of gravitation and electromagnetism treated in a unified manner in classical physics. 
 
 In section II, the basic equations needed for introducing the non perturbative guiding center transformation \cite{me0} are considered, and they are extended to relativistic regimes. Within such approach it will be possible to analytically describe the motion of a charged  (classical) particle in a general e.m. field. Some trivial solutions are shown in section III. These are the guiding particle solution which is minimally coupled with the magnetic field and the gyrating particle solution that describes a closed orbit trajectory spinning around a fixed guiding center. In section IV, the relativistic non perturbative guiding center description of single particle motion is described. Similar results are obtained in section V, by adopting the same lagrangian formalism used  for the magnetic force lines in \cite{cary1983}.  \\
Finally, in section VI, the particle dynamics are considered  with different metric tensors: from a flat space-time geometry ($M_4$) to a curved extended phase-space (position, velocity and time) geometry.  The novelty is that, instead of directly adding to the single particle lagrangian,  a term for taking into account the presence of e.m. fields, we prefer to add a HE-like lagrangian. Thus, the metric tensor could be determined through the variation of the HE action in extended phase-space. If the guiding center coordinates are employed, it will be possible to apply the KK mechanism \cite{kk1,kk2} with a geometry $\mathbb{R}^{3,1}\times S^1$ for the extended phase space so that the solution for the metric tensor is exactly the one proposed by KK.\\
The e.m. fluctuations are considered in Section VII and the analysis of solutions, which is the important issue studied in gyrokinetics, is considered here from a stochastic perspective. Thanks to such improvement on the gyrocenter transformation, if non relativistic energies are considered, it will be shown that the gyrocenter motion is fine described by the \emph{Schr\"odinger} equation. The possibility of reconciling general relativity with quantum mechanics is resolved by the fact that they describe different quantities, the general relativity  describes the guiding center, whereas quantum mechanics describe gyrocenters whose motion, with respect to guiding centers, is  also due to electromagnetic fluctuations.
 \\
The analysis is firstly done by adopting the Eulerian description of dynamical quantities. However, the final description of motion is done in the guiding center description. Even if the motion is independent on such choice, the privileged reference system here adopted is the guiding center one. In the appendix some details on the derivation of the KK mechanisms are reported following \cite{deluca}.
\section{Basic Equations}
A charged particle (charge $e$ and mass $m$) that moves in a \emph{given} e.m. field is classically described by the Lorentz's force law:
\begin{equation}
\label{lorentz}
\frac{d}{dt}\gamma_v v=\frac{e}{m}\left( E+v\times B \right),
\end{equation}
for the speed of light set to $1$.
The relativistic factor is $\gamma^{-1}_v=\sqrt{1-v^2}$ in the flat Minkowski spacetime. If $s$ is the \emph{proper time} or the \emph{world line coordinate}, then $\gamma^{-1}_v=\dot{s}$, where the dot is indicating the time derivative. In (\ref{lorentz}), $v=\dot x$ is the velocity. 
To obtain the solutions of (\ref{lorentz}), we use the \emph{newtonian} idea of a deterministic world. Following \cite{me0}, supposing to know the exact solutions of the motion, in such a way that it is possible to fix the velocity, $v$, for each point of the space (traced by the particle), $x$, at each time, $t$: $\dot x=v(t,x)$. The former equation indicates the \emph{pathline} in continuum mechanics \cite{Ottino}.
The motion will also depend on other quantities, \emph{e.g.} the initial energy $\varepsilon_0$, being $\varepsilon=\gamma_v+e \Phi/m$ ($\Phi$ is the electric potential), or the initial velocity, $v_0$. However, we treat such variables as constant parameters and, at the moment, they are not explicitly considered.
The total derivative with respect to time is:
\begin{equation}
\label{eq2}
\frac{d}{dt}\gamma_v v=\partial_t \gamma_v v+  v \cdot \nabla \gamma_v v= \partial_t \gamma_v v + \gamma^{-1}_v \nabla  \frac{\gamma^2_v v^2}{2} -  v \times \nabla \times \gamma_v v.
\end{equation}
Introducing the e.m. potentials, $\Phi$ and $A$, in (\ref{lorentz}) then the equation (\ref{eq2}) becomes
\begin{eqnarray}
\label{ohmslaw0}
\partial_t (\gamma_v v+e A/m)+\gamma^{-1}_v \nabla  \frac{\gamma^2_v v^2}{2}+(e/m)\nabla \Phi=\\
\nonumber
=v \times [ (e/m) B +\nabla \times v].
\end{eqnarray}
From the identities $\gamma^{-1}_v \nabla  \gamma^2_v v^2/2=\gamma^{-1}_v \nabla  \gamma^2_v /2=\nabla  \gamma_v$, it follows:
\begin{equation}
\partial_t (\gamma_v v+e A/m)+\nabla (\gamma_v +e \Phi/m)= v \times \nabla \times (\gamma_v v+e A/m).
\end{equation}
The latter equation can be suggestively read introducing the "canonical" e.m. fields $E_c=-(m/e) \nabla \varepsilon - (m/e) \partial_t p$ and $ B_c=(m/e) \nabla \times p$. In fact, $E_c$ and $B_c$ are said "canonical" because of the potentials, $\Phi_c= (m/e)\varepsilon$ and $A_c=(m/e)p$, that are the energy and momentum, \emph{i.e.} the canonical variables conjugated to time and position, respectively. Now, the equation (\ref{ohmslaw0}) is rewritten as
\begin{equation}
\label{ohmslaw}
E_c+v\times B_c=0,
\end{equation}
which means that in the reference frame that moves with the particle, $\dot x= v(t,x)$, the particle is seen always at rest. In fact, the resultant of forces vanishes in such \emph{co-moving} frame.
This is the \emph{free-fall reference frame} for electromagnetism and something similar to the \emph{equivalence principle} can also be stated here. The difference with the standard approach is that it has been adopted an \emph{eulerian} description of motion instead of the lagrangian one. The main differences between the two approaches are soon analyzed.  \\
\subsection{The Lagrangian and the Eulerian description of motion}
If the charge position at $t=0$ is known: $x(t=0)=x_0$, then the flow is represented by the map, $\Phi_t$, that determines the charge position at a later time:
\begin{equation}
\label{motion}
x(t)=\Phi_t(x_0),
\end{equation}
being $x_0=\Phi_{t=0}(x_0)$. In continuum mechanics, the former equation is simply named the \emph{motion}. Concerning the definition of Lagrangian vs Eulerian descriptions, we closely follow the textbook \cite{Ottino}. The Lagrangian velocity is defined to be $v=v(t,x_0)$, and it is referred to the charge $x_0$, that means the charge that initially was at $x_0$ (when $t=0$). The Eulerian velocity is defined to be $v=v(t,x)$, that gives the velocity when the particle $x_0$ passes through $x$ at time $t$. The same is true for any quantities, \emph{e.g.} $\mathit{O}$ can be expressed in Lagrangian description, then $\mathit{O}=\mathit{O}(t,x_0)$ and the particular charge $x_0$ is followed in its time evolution, otherwise, in eulerian description, $\mathit{O}=\mathit{O}(t,x)$, and $\mathit{O}$ refers to the charge $x_0$ when it passes through $x$ at time $t$.  The time derivative is often called the  \emph{material} derivative:  $\dot{\mathit{O}}=\partial_t \mathit{O} |_{x_0}=\partial_t \mathit{O} |_x+\dot{x}\cdot \nabla \mathit{O}=\partial_t \mathit{O}+ v \cdot \nabla \mathit{O}$, for the chain rule. What is important and heavily used in the present work is the fact that the computation of the acceleration, $a=\dot{v}$, at $(t,x)$ can be done without solving the motion first. This only enables  the knowledge of $v=v(t,x)$ and not of $x=\Phi_t(x_0)$:
\begin{equation}
a=\partial_t v+ v \cdot \nabla v.
\end{equation}
\subsubsection{Note on lagrangian mechanics}
The non relativistic \emph{Lorentz' force law} is the same of equation (\ref{lorentz}) with the substitution $\gamma_v=1$. It is quite simple to obtain such force from the \emph{Euler-Lagrange} (EL) equations:
\begin{equation}
\frac{d}{dt}\nabla_{\dot{x}} L_\mathrm{nr}-\nabla L_\mathrm{nr}=0,
\end{equation} 
where the non relatvistic lagrangian, $L_\mathrm{nr}=L_\mathrm{nr}(t,x,\dot{x})$, is
 \begin{equation}
 \label{L0}
 L_\mathrm{nr}(t,x,\dot{x})=\frac{\dot{x}^2}{2}+(e/m)\dot{x}\cdot A(t,x)-(e/m)\Phi(t,x).
 \end{equation}
  It is remarkable that the EL equations can be obtained from a variational principle, i.e. the $\emph{Hamilton's principle}$. If the action is defined to be
 \begin{equation}
 S_\mathrm{nr}=\int^{t_\mathrm{out}}_{t_\mathrm{in}} L_\mathrm{nr}\, dt. 
 \end{equation}
 being $t_\mathrm{out}$ and $t_\mathrm{in}$ two different instants of time, then it is possible to associate the EL equations with an extremal of the action. If \emph{all} the trajectories are considered, from $t_\mathrm{in}$ and $t_\mathrm{out}$, there are \emph{some} of those trajectories for which the action is at an extremal. Let's take a trajectory of motion, $x=X(t)$ that passes in $X_\mathrm{in}$ at $t_\mathrm{in}$ and in $X_\mathrm{out}$ at $t_\mathrm{out}$. Such trajectory is the so-called trajectory of motion because it is solution of $\ddot X=(e/m)(E+\dot X\times B)$. Starting from such trajectory it is possible to consider all the other trajectories that are parametrically written at each instant of time, $t$, as
 \begin{eqnarray}
\label{transf}
x=X+\rho(t,X,\alpha)\\
\nonumber
\dot{x}=\dot{X}+\dot{\rho}(t,X,\alpha),
\end{eqnarray}
  where $\alpha=(\alpha_1,\alpha_2,\alpha_3)$ could vary on a three dimensional domain. It is useful to ask for the following property: if $\alpha_2$ goes to zero, then also $\rho$ goes to zero and the considered trajectory collapses on the trajectory of motion, $x$ goes to $X$ (and $\dot x$ goes to $\dot X$). With respect to the standard approach we are considering all the trajectories, not only the one starting from  $X_\mathrm{in}$ at $t_\mathrm{in}$ to $X_\mathrm{out}$ at $t_\mathrm{out}$. Such difference causes the following consequence. The variation of the action ( with respect to the parametric space), $\delta S_\mathrm{nr}$ is always given by
  \begin{equation}
  \delta S_\mathrm{nr}=\rho \cdot \nabla_{\dot{x}} L_\mathrm{nr}|_{t_\mathrm{in}}^{t_\mathrm{out}}-\int^{t_\mathrm{out}}_{t_\mathrm{in}}\left[\frac{d}{dt}\nabla_{\dot{x}} L_\mathrm{nr}-\nabla L_\mathrm{nr}\right] \cdot \rho \, dt,    
  \end{equation}
  but now the EL equations doesn't ensure that $\delta S_\mathrm{nr}=0$ because of the term $\rho \cdot \nabla_{\dot{x}} L_\mathrm{nr}$ which can be different from zero. The common practice is to consider $\rho=0$ at $t=t_\mathrm{in}$ like at $t=t_\mathrm{out}$. However, this is not necessary. You can also consider all the trajectories with $\rho \cdot \nabla_{\dot{x}} L_\mathrm{nr}=0$ but $\rho(t_\mathrm{in})\neq0$ and  $\rho(t_\mathrm{out})\neq0$, and, again, the result is that the force law corresponds to the vanishing of the first variation of the action. In this case there are many (infinite) trajectories for which the EL equations (\emph{i.e.} Lorentz' force law) are satisfied, even if the coordinates $X,\dot{X}$ are always describing the unique trajectory that starts from $X_\mathrm{in}$ at $t_\mathrm{in}$ to reach  $X_\mathrm{out}$ at $t_\mathrm{out}$. In such case, both the EL equations and the Hamilton's principle are satisfied, even if $\alpha_2\neq0$. The reason for noting such difference with respect to the standard approach is quite unimportant unless there is something, like an indetermination principle or some non-locality properties,  that doesn't allow to exactly known where the particle is at $t_\mathrm{in}$  and at $t_\mathrm{out}$. We will see in section VI that the present theory is non-local and the latter extended approach to the variational description is useful. Moreover, in section VII, it is shown that $\alpha_2 \neq 0$ \emph{almost always} and the \emph{classical trajectory} (with $\alpha_2=0$) is ruled out by electromagnetic fluctuations.
 \subsubsection{The non relativistic case}
 In (\ref{L0}), it is possible to substitute the potentials, that are fields, \emph{i.e.} functions of time and position, with other physically meaningful fields. For an \emph{arbitrary} velocity field, $V(t,x)$, it is possible to define $\mathcal{E}(t,x)=V^2/2+(e/m)\Phi(t,x)$ and $P(t,x)=V +(e/m)A$. Also $\mathcal{E}=\mathcal{E}(t,x)$ and $P=P(t,x)$ are fields. The Lagrangian becomes 
 \begin{eqnarray}
 \label{L1}
 \nonumber
&&L_\mathrm{nr}=\frac{\dot{x}^2}{2}+\dot{x}\cdot(P-V)(t,x) -(\mathcal{E}-V^2/2)(t,x)=\\
&&=\frac{(\dot{x}-V)^2}{2}+P \cdot \dot{x}-\mathcal{E}.
 \end{eqnarray}
 The momentum, $p \equiv \nabla_{\dot{x}}L$, is $p=\dot{x}-V+P$.\\ It is worth noticing that the arbitrariness of $V$ is very important. Behind such arbitrariness there is the \emph{relativity principle}. In fact, the presence of V can be seen as a particular choice of the reference frame in the space of velocities and, therefore, does not affect the dynamics.  
 It is not important if the observer of an experiment moves with an arbitrary velocity $V$, the physics described by the experiment remains the same because the lagrangian has the same value, being a scalar. The property of the lagrangian of being a scalar \emph{is} the relativity principle and it will be very useful in next sections. \\
 Now, it is easy to recognize two different descriptions of the same motion, the Lagrangian description, which  is almost adopted, is when $V=0$ and the \emph{Lorentz's force law} is recovered. Whilst, for $\dot{x}=V(t,x)$, the description is said Eulerian.\\
 Concerning the \emph{Euler-Lagrange} (EL) equations, they are computed:
 \begin{eqnarray}
 &&\frac{d}{dt}(\dot{x}-V)=-\partial_t P -\nabla \mathcal{E} + \dot{x} \times \nabla \times P+\\
 \nonumber
 &&-(\dot{x}-V)\times \nabla \times V-(\dot{x}-V)\cdot \nabla V. 
 \end{eqnarray}
 Introducing the \emph{canonical} e.m. fields $(e/m)E_c=-\partial_t P -\nabla \mathcal{E}$ and $(e/m)B_c= \nabla \times P$, the former can be rewritten as
  \begin{eqnarray}
   \label{el0}
&& \frac{d}{dt}(\dot{x}-V)=(e/m)(E_c+\dot{x} \times B_c)+\\
\nonumber
&&-(\dot{x}-V)\times \nabla \times V-(\dot{x}-V)\cdot \nabla V.
 \end{eqnarray}
 It is now evident that a solution of motion is when $\dot{x}=V$ and $V$ is solution of equation (\ref{ohmslaw}): 
 \begin{equation}
 \label{ohmslawV}
 E_c+V \times B_c=0,
 \end{equation}
 which seems, only apparently, an algebraic equation. \\
 In dynamical systems or continuum mechanics, given the \emph{eulerian} velocity field $V=V(t,x)$, \emph{i.e.} the velocity of the charge when it passes at $x$ at time $t$, the problem is to find the \emph{particle path}, integrating the equation $\dot{x}=V(t,x)$. Differently, here the eulerian velocity is not given and we have to solve equation (\ref{ohmslaw}) to obtain the velocity field, eventually for trying to integrate the motion (which is not our first interest). We concisely refer to equation (\ref{ohmslaw}) as the \emph{velocity law} because it can be found for every e.m. fields and for every charge if the eulerian description is adopted. Moreover, with respect to an observer that moves with the Eulerian velocity $V(t,x)$, from equation (\ref{el0}), the electric field is $E_c+V\times B_c$ which is null so that the charge is kept at rest. With respect to an observer co-moving with the laboratory, the e.m. fields can be measured to be $E$ and $B$ whilst the charge is seen to move following the Lorentz's force law. \\
 Another interpretation of the same equation, is the following. Suppose to realize, in a laboratory, the  electric field, $E_c$, and the magnetic field, $B_c$,  and to be able to move the charge in $X$ at $t$ in such a way that its velocity is described by $V(t,X)$. Then, the Lorentz' force on the particle vanishes, being $E_c+V \times B_c=0$. Without a force on the charge, it is possible to consider the charge velocity preserved as in an  inertial reference frame. The problem is that the velocity is not constant and the trajectory is not straight as for a global inertial reference frame. This is exactly what occurs if the reference frame is considered inertial only locally as it happens when an equivalence principle is considered. \\
  It is worth noticing that there is an interesting similarity between such equations of motion and the \emph{ideal Ohm's law} encountered in magneto-hydro dynamics (MHD). In MHD, the ideal Ohm's law is below written: $E+V_p\times B=0$, being $V_p$ the plasma eulerian velocity. Thus, even if the context is different, the solutions are similarly classified ( see also \cite{pegoraro2015} for the true relativistic Ohm's law ).
  \\
  If $B_c\neq0$, it is possible to rewrite $ E_c+V \times B_c=0$ as
\begin{equation}
\label{MHDsol}
V=V_b b + \frac{E_c \times B_c}{B^2_c}, 
\end{equation}
where $b$ is the unit vector in the direction of the canonical magnetic field, $B_c=|B_c| b$, and $E_c \times B_c/B^2_c$ is the $E\times B$-like \emph{drift} velocity.
In plasma physics, it is interesting to study the case corresponding to the \emph{gyro-kinetic} ordering  that neglects the $E\times B$-like drift.\\
Last but not least,  in the Eulerian description the lagrangian in (\ref{L1}) corresponds to the Poicar{\'e}-Cartan form, which linearly depends on the velocity:
 \begin{equation}
 \label{PClagr}
 L_\mathcal{PC}=P(t,x)\cdot \dot x-\mathcal{E}(t,x).
 \end{equation}
    \subsubsection{The relativistic case}
    When relativistic energies are considered it is important to give a covariant description. In this section the spacetime is considered \emph{Minkowskian} (flat geometry) with signature $\eta_{\alpha \beta}=\mbox{diag}(1,-1,-1,-1)$.
Let's start from the scalar \emph{Lagrangian}:
\begin{equation}
\label{lag1}
L=-1+(e/m) (A\cdot u -\Phi \sqrt{1+u^2}),
\end{equation}
being $\sqrt{1+u^2}=\gamma_v$. We indicate with the \emph{prime} the derivative with respect to the world line coordinate, $s$, so that $u=x^\prime$ is the relativistic velocity. The lagrangian (\ref{lag1}) is the sum of two effects, the free single particle lagrangian is $L_\mathrm{free}=-1$ while the lagrangian expressing the interaction between matter and the e.m. field is $L_\mathrm{ime}=(e/m) (A\cdot u -\Phi \sqrt{1+u^2})$.  Adopting the \emph{summation convention} and for $u^\alpha u_\alpha=1$ with $\alpha=0,1,2,3$, the lagrangian can be re-written in the familiar form
\begin{equation}
\label{relLan}
L=- u^\alpha (u_\alpha + e A_\alpha/m),
\end{equation}
being $A_0=\Phi$ the electric potential. Explicitly, we have assumed that the \emph{contravariant} velocity is $u^\alpha=(\gamma_v,\gamma_v v)$, while the \emph{covariant} velocity is  obtained from the product $u_\alpha=\eta_{\alpha \beta} u^\beta$, that gives $u_\alpha=(\gamma_v,-\gamma_v v)$.\\
From  $1=1/2+u^\alpha u_\alpha/2$,  an equivalent lagrangian can be written:
\begin{equation}
\label{covariantL}
L=-\frac{u^\alpha u_\alpha}{2}-\frac{e}{m} u^ \alpha A_\alpha-\frac{1}{2}.
\end{equation}
It is worth to note that such lagrangian is very similar to the non relativistic one, $L_\mathrm{nr}=v^2/2+(e/m)(v\cdot A-\Phi)$; if $u \to v$ then the difference is only due to the energy at rest, which is absent in $L_\mathrm{nr}$. 

Now, for an arbitrary four co-variant velocity field, $U_\alpha=U_\alpha(x^\beta)$ it is possible to define a co-vector field, $P_\alpha=P_\alpha(x^\beta)=U_\alpha+(e/m)A_\alpha$.  The Lagrangian becomes
\begin{equation}
L=-\frac{u^\alpha u_\alpha}{2}-u^ \alpha (P_\alpha-U_\alpha)-\frac{1}{2}.
\end{equation}
The four co-momentum are
\begin{equation}
p_\alpha=-\frac{\partial L}{\partial u^\alpha}=u_\alpha+P_\alpha-U_\alpha.
\end{equation} 
The EL equations are simply
\begin{equation}
\frac{d}{ds} p_\alpha=u^\beta\partial_\alpha (P_\beta-U_\beta),
\end{equation}
that, finally, can be written as:
\begin{equation}
\frac{d}{ds} (u_\alpha-U_\alpha)=u^\beta(\partial_\alpha P_\beta-\partial_\beta P_\alpha)-u^\beta \partial_\alpha U_\beta.
\end{equation}
As before, if $U_\alpha=0$ the equations of motion give the covariant Lorentz's force law, and a Lagrangian description is preferable.  However, if $u_\alpha=U_\alpha(x^\beta)$, then the description is Eulerian. In the Eulerian description, the Eulerian four velocity satisfy the equation (\ref{ohmslaw}) because $u^\alpha u_\alpha=1$ and $u^\beta \partial_\alpha u_\beta=0$. The velocity law can be written in co-variant form as
\begin{equation}
\label{idlohm}
u^\beta \omega_{\alpha \beta}=0,
\end{equation}
being $\omega_{\alpha \beta}=\partial_\alpha P_\beta-\partial_\beta P_\alpha$, known as the \emph{Lagrange tensor}.\\
The \emph{canonical Maxwell} tensor, $F_{c \, \alpha \beta}$ is proportional to the Lagrange tensor, $\omega_{\alpha \beta}$:
\begin{equation}
\label{canMaxwtens}
 (e/m)F_{c \,\alpha \beta}\equiv \omega_{\alpha \beta} = \partial_\alpha  P_\beta - \partial_\beta  P_\alpha. 
\end{equation}  
  Thus, equation (\ref{ohmslaw}) is found when $\alpha=1,2,3$; whilst, if $\alpha=0$ then  
\begin{equation}
\gamma_v v \cdot (- \nabla \mathcal{E} - \partial_t P)=0, 
\end{equation}
which means that $E_c$ is transversal to $v$, so that it doesn't contribute to the energy variation (for this reason, in \cite{me0}, $E_c$ was indicated as $E_t$). \\
Even if we have already obtained the covariant equations, it is instructive to derive the same equations (\ref{idlohm}) directly from the most simple lagrangian:
$L=-u^\alpha p_\alpha(x^\beta)$, which is the same of  (\ref{relLan}) but now the covariant momentum, $p_\alpha=p_\alpha(x^\beta)$, is only function of the spacetime coordinates and it doesn't depend on the (relativistic) velocity (its one-form is the \emph{Poincar{\'e}-Cartan} form). For such lagrangian, the four momentum is
\begin{equation}
p_\alpha(x^\beta)\equiv-\frac{\partial}{\partial u^\alpha} L
\end{equation}
and 
the EL equations are:
\begin{equation}
u^\beta \partial_\beta p_\alpha = u^\beta\partial_\alpha p_\beta,
\end{equation}
 being $p^\prime_\alpha=u^\beta \partial_\beta p_\alpha$. 
The former is exactly the equation (\ref{idlohm}).
 \section{Solutions of the velocity law}
There are some simple solutions of the equation (\ref{idlohm}). 
The trivial solution, $\omega_{\alpha \beta}=0$, results to be very important. Another simple solution is $\omega_{\alpha \beta}=\epsilon_{ \alpha \beta \gamma \delta} k^\gamma u^\delta$, where $\epsilon_{\alpha \beta \gamma  \delta}$ is the \emph{Levi-Civita} symbol ($\epsilon_{0123}=1$) and $k^\gamma$ is the \emph{wave number} four-vector. Also this solution is trivial because the \emph{Levi-Civita} symbol is totally anti-symmetric, so that  $u^\beta \epsilon_{ \alpha \beta \gamma \delta} k^\gamma u^\delta =0$ due to the symmetry $\beta \leftrightarrow \delta$.\\
From equation (\ref{MHDsol}) it is possible to classify the solutions of the velocity law depending on i) $B_c=0$ and $E_c=0$, ii) $B_c\neq 0$ and $E_c=0$ and iii) $B_c \neq 0$ and $E_c \neq 0$. They are called, gyrating particle solution, guiding center solution in gyrokinetics ordering and guiding center solution in MHD ordering, respectively.  
\subsection{Relativistic guiding particle solution}
Let's  start with the analysis of the following solution: $\omega_{\alpha \beta}=\epsilon_{ \alpha \beta \gamma \delta} k^\gamma u^\delta$ and consider the case $k^0=1/\lambda$ and $k=0$. Now,  $\omega_{\alpha \beta}=u^\delta \epsilon_{\delta \alpha \beta 0}/ \lambda$. Thus, only the spatial components survive:
\begin{equation}
\omega_{i j}=\frac{u^k}{\lambda}\epsilon_{i j k}  \qquad \mbox{   with  } i,j,k=1,2,3.
\end{equation}
Multiplying for $\epsilon^{i j l}$ both sides of the latter equation, and using the equivalence $\epsilon_{ijk}\epsilon^{ijl}=2\delta^k_l$, then
\begin{equation}
\frac{u^l}{\lambda}=\frac{\epsilon^{ijl}}{2} \omega_{ij}=\epsilon^{ijl} \partial_i p_j ,
\end{equation}
which is the $l$ component of the \emph{fundamental equation}:
\begin{equation}
\label{claudio1}
\frac{u}{\lambda}=\frac{e}{m}B+\nabla \times u,
\end{equation}
 as in the non-relativistic case \cite{me0}.  \\
  The latter equation is said fundamental because its solution gives the answer for many problems encountered in plasma physics and/or electrodynamics. At first, if $\lambda \to \infty$ then $B=-(m/e) \nabla \times u$ and the velocity becomes strictly related to the vector potential: the problem is to find a vector potential from a given magnetic field. This kind of solution will be called the \emph{gyrating particle} solution. Secondly,  if $(e/m) \to 0$ then $u=\lambda \nabla \times u$, which is recognized as the \emph{force free equation} \cite{lust,chandra} that denotes the \emph{Beltrami field} \cite{beltrami}. In \cite{me0},  equation (\ref{claudio1}) is treated as the non-homegeneus version of the \emph{force free} equation. Finally, the \emph{guiding particle solution} is obtained when the vorticity, $\nabla \times u$, is small: $\nabla \times u \sim 0$. In this case, the velocity is mostly parallel to the magnetic field $u \sim (e/m) \lambda B$, and the vorticity gives the drift velocity \cite{me0}: $v_D =\lambda \nabla \times u$. In \cite{mahajan}, the same equation is part of a system of equations where the equation (\ref{claudio1}) is coupled with  another similar equation that describes the magnetic field. Such system of equations is used for describing interesting diamagnetic structures in plasmas.\\
   Nevertheless, written in the latter form, something unusual appears. In fact, the \emph{Hamilton-Jacobi} solutions, that are \emph{classical} solutions, are obtained  setting $p=\nabla S$, where $S$ is the principal \emph{Hamilton} function\footnote{ In the present case $S=W-\varepsilon t$, where $W$ is the \emph{Hamilton's characteristic function} and $\nabla S= \nabla W$.}. In our case, $p$ is not a gradient of a function, otherwise its \emph{curl} should vanish.  We have already defined the \emph{canonical} magnetic field exactly as the \emph{curl} of $(m/e)p$. This means that classical solutions have $B_c=0$ and so, we are inspecting \emph{non classical} solutions with $B_c\neq 0$.\\
 Together with (\ref{claudio1}), there is also the condition $\omega_{0i}=0$, which means:
\begin{equation}
\partial_t p +\nabla \varepsilon=0,
\end{equation}
as it should in the gyrokinetic-like ordering ($E_c=0$).
Such equation, already studied in \cite{me0}, is particularly important when $u \sim \lambda eB/m$. In this case it is better to indicate $u$ with $U$ and refer to it as the \emph{guiding particle relativistic velocity}. 
The reason is that it describes the motion of a particle, with null magnetic moment  that proceeds mostly parallel to the magnetic field with a drift velocity $\lambda \nabla \times U$.
 For a generic magnetic field, it is possible
 to obtain a perturbative solution ordered in power of $\lambda$ so that the $0^{th}$ order approximation is
 \begin{equation}
 U^{(0)}=\frac{e}{m} \lambda B.
 \end{equation}
 The leading order approximation is
\begin{equation}
U^{(1)}= \lambda  \frac{e}{m} B+\lambda \nabla \times U^{(0)}.
\end{equation}
If $\lambda=(m/e)u_\|/|B|$, then the former is the familiar guiding center (relativistic) velocity (for null magnetic moment) at leading order:
\begin{equation}
U^{(1)}= u_\| b_{(0)}+ \frac{u_\|}{(e/m)|B|} \nabla \times u_\| b_{(0)},
\end{equation}
with $B=|B| b_{(0)}$.
In \cite{me0}, an exact solution of (\ref{claudio1}) is obtained when the magnetic field is axisymmetric as it happens in many interesting circumstances. In such case, a common representation of $B$ is
\begin{equation}
\label{magnf}
B=\nabla \psi_p \times \nabla \phi + F\nabla \phi,
\end{equation} 
where $\phi$ is the toroidal angle, $\psi_p$ is the {poloidal magnetic flux surface} and $F/R$ is the toroidal component of the magnetic field ($\nabla \phi=e_\phi/R$ with $e_\phi$, the unit vector in the toroidal direction, and $R$ the radial distance from the axis).  The guiding particle velocity solution of  (\ref{claudio1}) in an axisymmetric magnetic field like (\ref{magnf}) is \cite{me0}
\begin{equation}
\label{VaxiB}
U= \lambda \frac{e}{m}\nabla \mathcal{P}_\phi \times \nabla \phi + \frac{e}{m} (\mathcal{P}_\phi-\psi_p )\nabla \phi
\end{equation}
with the guiding particle toroidal momentum, $\mathcal{P}_\phi$ (in magnetic flux unit), satisfying the following equation:
\begin{equation}
\lambda R^2 \nabla \cdot \frac{\lambda \nabla \mathcal{P}_\phi}{R^2}+\mathcal{P}_\phi=0,
\end{equation} 
if $\lambda=-\psi_p/F$.
The latter equation, that can be written as an eigenvalue equation for the \emph{Shafranov operator}, was already obtained but wrongly written in \cite{me0} (see \cite{me2} for details).
\subsection{Velocity law solutions in MHD-like orderings}
Previously, we have analyzed the following  solution of  the velocity law: $\omega_{\alpha \beta}=\epsilon_{ \alpha \beta \gamma \delta} k^\gamma u^\delta$, with $k^\gamma$ the \emph{time-like} four-vector: $k^0=1/\lambda$ and $k=0$. We have noticed that from this choice it follows that $E_c=0$, which is said the gyro-kinetic-like ordering.  Now we want to consider the case where $k^\gamma$ is the \emph{space-like} four vector $(0,k)$. In such case, $\omega_{\alpha \beta}=\epsilon_{ \alpha \beta i \delta} k^i u^\delta=\epsilon_{ \alpha \beta i 0} k^i u^0+\epsilon_{ \alpha \beta i j} k^i u^j$. 
The component of $\omega_{\alpha \beta}$ are
\begin{equation}
\omega_{0k}=\epsilon_{ 0 k i j} k^i u^j= (k \times u)_k
\end{equation}
and
\begin{equation}
\omega_{kj}=\epsilon_{ k j i 0} k^i u^0.
\end{equation}
That can be written  in  vectorial form as
\begin{equation}
E_c=(m/e)\gamma_v k \times  v
\end{equation}
and
\begin{equation}
B_c=(m/e)\gamma_v k,
\end{equation} 
being $\omega_{\alpha \beta}=(e/m) \mathcal{F}_{c \, \alpha \beta}$.
In such case it is the wave number, and not $v$, that is parallel to $B_c$.
The solution for $v$ is the same of  (\ref{MHDsol}) but contrary to before  the particles don't follow trajectories close to the magnetic field lines because of the presence of the electric field $E_c$. This is what happens in the MHD-like ordering. Thus,  we can easily distinguish the MHD-like from the gyrokinetic ordering giving to $k^\sigma$ the character of a space-like or time-like four-vector, respectively. The same conclusion can also be done if the \emph{ideal Ohm's law} is considered instead of the velocity law of equation (\ref{ohmslaw}).
\subsection{Gyrating particle solution}
The trivial solution of (\ref{idlohm}) is $\omega_{\alpha \beta} =0$. In such case the canonical fields are null: $\mathcal{F}_{c \, \alpha \beta}=0$, or $E_c=B_c=0$. \\
If $B_c=0$, then $(e/m)B+\nabla \times u=0$. Now, it is possible to choose a very particular vector potential: $A=-(m/e)u + \nabla g$, being $g$ a \emph{gauge} function. Moreover, the function $g$ is also seen to be proportional to the \emph{principal Hamilton's function}, $S$, which is an \emph{action}. Indeed, the canonical momentum is $p=u+(e/m)A=(e/m)\nabla g$. If $g=(m/e)S$, then $p=\nabla S$ and we have just set the initial condition for finding the \emph{classical} solution in the \emph{Hamilton-Jacobi} method. This is also consistent with $E_c=0$, that means that $\nabla u^0 + (e/m)\nabla \Phi + (e/m)\partial_t \nabla g=0$. If $\varepsilon=u^0 + (e/m) \Phi= -(e/m)\partial_t g$, then we found the other \emph{Hamilton-Jacobi} equation: $\varepsilon+\partial_t S=0$, being $S=(e/m)g$. \\ 
\subsubsection{Zitterbewegung}
The gyrating particle solution is the most important solution.  The reason will only be clear in Section VII, but it is possible to notice some interesting properties also here if the gauge function is settled to $g=(m/e)^2\mu\gamma$ (such choice will soon be defined as the \emph{guiding center gauge function}). The four-vector co-momentum is $p_\alpha=-(m/e)\mu\partial_\alpha \gamma$. Explicitly, $p=(m/e)\mu \nabla \gamma$ and $\varepsilon=-(m/e) \mu \partial_t \gamma$.\\
 Such solution allows to compute the Lagrangian, whose value is:
$ L=v\cdot p-\varepsilon=(m/e)\mu( \partial_t \gamma+v\cdot \nabla \gamma)=(m/e)\mu \dot{\gamma}$. If $\gamma$ is the gyro-phase, the conjugate coordinate, $\mu$, is the magnetic moment. Thus, the gyrophase comes to be proportional to the action. Such remark was already considered by Varma \cite{varma}, who firstly recognizes the importance of identifying the gyrophase with the action in another context: \emph{path integral} formulation of quantum mechanics for discovering quantum effects on macro-scale dynamics. In the present work, it is not possible to sufficiently stress why we should consider the gyrophase an action coordinate and it will be considered elsewhere together with the Varma's idea.  
However, another surprising correspondence with quantum mechanics is here described.   Concerning the non relativistic energy of the charge,
\begin{equation}
\label{nrenergy}
\varepsilon=\frac{[p-(e/m)A]^2}{2}+\frac{e}{m}\Phi.
\end{equation}
it can be re-written in an interesting way once the phase function, $\psi_\mathrm{zbw}=e^{-i\gamma}$, is considered. Here, $\mathrm{zbw}$ stands for \emph{Zitterbewegung} \cite{Zitter,huang,barut,hestenes}. From the derivatives of the phase function
\begin{equation}
\nabla \psi_\mathrm{zbw}=-i\psi_\mathrm{zbw}\nabla\gamma \qquad \mbox{ and } \qquad  \partial_t \psi_\mathrm{zbw}=-i\psi_\mathrm{zbw}\partial_t\gamma.
\end{equation}
Thus, the momentum and the energy can be written as
\begin{eqnarray}
&&p=i(m/e)\mu\psi^\star_\mathrm{zbw}\nabla\psi_\mathrm{zbw},\\
&&\varepsilon=-i(m/e)\mu\psi^\star_\mathrm{zbw}\partial_t \psi_\mathrm{zbw},
\end{eqnarray}
where $\psi^\star_\mathrm{zbw}$ indicates the complex conjugate of $\psi_\mathrm{zbw}$, which, in this case, it is also the inverse: $\psi^\star_\mathrm{zbw}=\psi^{-1}_\mathrm{zbw}$. The energy is computed:
\begin{eqnarray*}
&&-i(m/e)\mu\psi^\star_\mathrm{zbw}\partial_t \psi_\mathrm{zbw}=\\
&&\frac{1}{2}\left[i(m/e)\mu\psi^\star_\mathrm{zbw} \nabla \psi_\mathrm{zbw}-(e/m)A\right]\cdot\\
&&\cdot \left[ -i(m/e)\mu\psi_\mathrm{zbw} \nabla \psi^\star_\mathrm{zbw}-(e/m)A\right]+\frac{e}{m}\Phi,
\end{eqnarray*}
 being $p$ and $\varepsilon$ real quantities.
The former expression is written as
\begin{eqnarray*}
&&-i(m/e)\mu\psi^\star_\mathrm{zbw}\partial_t \psi_\mathrm{zbw}=\\
&&\psi^\star_\mathrm{zbw}\frac{[i(m/e)\mu \nabla-(e/m)A]^2}{2} \psi_\mathrm{zbw}+\\
&&+\frac{i}{2}\nabla\cdot[i(m/e)\mu\psi^\star_\mathrm{zbw} \nabla \psi_\mathrm{zbw}-(e/m)A]+\frac{e}{m}\Phi.
\end{eqnarray*}
The divergency term is $\nabla \cdot v$ and it is null for an incompressible charge, neither created nor destroyed. A \emph{Schr\"odinger}-like equation is obtained:
 \begin{eqnarray}
 \label{schlike}
 &&-i(m/e)\mu\partial_t \psi_\mathrm{zbw}=\\
 \nonumber
 &&\frac{[i(m/e)\mu \nabla-(e/m)A]^2}{2} \psi_\mathrm{zbw}+\frac{e}{m}\Phi \psi_\mathrm{zbw}.
 \end{eqnarray}
 The operator, of the former partial differential equation, is exactly the same of the Schr\"odinger equation if $e<0$, as for an electron, and $\mu=|e|\hbar/m^2$.
From totally classical assumptions, the expression of the energy for a charged particle together with the constrain that $\nabla \cdot v=0$, give rise to a wave equation. Thus, for a gyrating particle solution it is possible to describe a classical solution of motion through a Schr\"odinger equation applied to a phase function like $ \psi_\mathrm{zbw}$. It is interesting that there is a  set of solutions where both the Schr\"odinger-like equation and the Lorentz's force law are verified. What is important is that determinism is preserved by the fact that $|\psi_\mathrm{zbw}|=1$, which means, in the \emph{Born} interpretation of $|\psi_\mathrm{zbw}|^2$, that the probability of finding the particle in the state represented by $\psi_\mathrm{zbw}$ is \emph{almost certain} (only if you know the initial gyrophase). 

\subsubsection{Magnetic flux linked to closed loops}
For the gyrating particle solutions, it is possible to take the following representation for the magnetic field: $B=\nabla \Psi \times \nabla \gamma$, which is commonly called \emph{Clebsh} representation. $\Psi$ and $\gamma$ are said \emph{Clebsh potentials}. Topologically, it is possible to choose $\gamma \in S^1$, in such a way that, in this case, it is considered the \emph{gyro-phase}. The variable $\Psi$ is the magnetic flux linked to the closed loop traced by $\gamma$.   $\nabla \Psi$ is orthogonal to $\nabla \gamma$, in such a way that $B$ doesn't depend on $\gamma$. The particle, that travels along the closed loop of curvilinear coordinate $\gamma$, always feels the same orthogonal magnetic force.  This happens for the particular representation of the magnetic field, not because the magnetic field is straight and uniform. For this reason, such representation is also known as the straight field line representation.
The motion of the gyrating particle is expressed by:
\begin{eqnarray*}
x=X+\rho(\gamma)\\
u=U+\nu(\gamma)=\nu(\gamma),
\end{eqnarray*}
where we have set $U=0$, so that the guiding center, $X$, is fixed. 
The gyrating loop motion has been described in \cite{me0} setting 
\begin{equation}
\label{closedloop}
\nu=\rho \times \Omega,
\end{equation}
where $\Omega$ is the relativistic angular frequency, that depends on the position of the particle. It is possible to choose  a local tern of orthogonal unit vectors: $e_\gamma \cdot e_\rho \times b=1$, where $\rho=|\rho| e_\rho$, $\Omega\cdot b=\gamma^\prime$ and $\nu=|\rho| \gamma^\prime e_\gamma$.\\
 It is worth noticing that $|\rho|$ becomes a conserved quantity if (\ref{closedloop}) is allowed. In fact, the world line derivative of $\rho^2/2$ is $\rho \cdot \rho^\prime=\rho \cdot \rho \times \Omega=0$. Thus, $|\rho|$ doesn't depend on $\gamma$. However, $|\rho|$ depends on the magnetic flux linked to the closed orbit. In other words, distances are now measured in terms of $\Psi$. Moreover, also the time of one revolution depends on $\Psi$, so that it can be considered a good time-like coordinate. \\
From (\ref{closedloop}), the closed trajectory lives on the surface of a sphere, $S^2$, of radius $|\rho|$. However, if other coordinates are used then the same particle is seen to move on a helicoidal trajectory. The circle $S^1$ is both the representation of the particle orbit, but also the description of the gyrating motion in the guiding center coordinates. What is important is that there is no difference from the point of sight of the particle. The particle moves in a circle, ignoring the rest of the world because it can only feel the effect of the Lorentz' force with the same magnetic field intensity. If the charge is described in the guiding center reference frame, then the magnetic field is always orthogonal to its direction of motion. In a certain sense, it is similar to a massive body in a gravitational field: the massive body moves straight along the geodesic but the spacetime is curved due to the presence of a gravitational field and the body is seen from an observer, \emph{e.g.} to fall versus another massive body. On the same footing, a charged particle moves circularly but the spacetime is measured in units of magnetic field and if such magnetic field is non uniform then the charge is seen from an observer with a relative relativistic velocity $-U\neq 0$, \emph{e.g.} to follow a helicoidal trajectory. In the forthcoming sections, the guiding center description of motion will be described, when both $U$ and $\rho$ are not vanishing, and it will be shown how the electromagnetism can be described within the formalism of general relativity. 
 \section{Guiding center description}
  Here,  another interesting and important description of motion is considered,   with respect to the Eulerian and the Lagrangian descriptions: the guiding center description of motion which is neither Lagrangian nor Eulerian. Such possibility arises if the velocity of the charge is written as $\dot x=V+\sigma$. Now, the motion is not Eulerian, because $V\neq \dot x$, and it doesn't express the velocity of the charge at a given time and position. $V$ expresses the velocity field that is used as a system of reference for velocities measured by $\sigma=\dot{x}-V$. Such description is not Lagrangian, because $V\neq 0$, now. However, if the vector field $\sigma$ is defined on the same domain of $V$, then re-naming $\tilde V=V+\sigma$,  the description can be done in the Lagrangian or Eulerian way, as previously done. Instead, for the guiding center description we do something different, now $\sigma$ depends on a new variable or parameter, $\gamma \in S^1$, which it can be identified with the gyro-phase. In practice, $\gamma$ must live in a different domain from the one where the e.m. fields are defined and, thanks to Faraday,  the e.m. field only varies on space-time. Such new variable is always part of the whole phase-space. The non perturbative guiding center transformation is the transformation from $(t,x,v)$  to $(t,X,\gamma,\mu,\varepsilon)$, where $\mu$ is the magnetic moment, that will be defined later on, whilst $\varepsilon$ is the energy of the charge. The transformation is implicitly written as
  \begin{eqnarray}
\label{gctransf}
x=X+\rho(t,X,\gamma;\mu,\varepsilon)\\
\nonumber
v=V(t,X;\mu,\varepsilon)+\sigma(t,X,\gamma;\mu,\varepsilon).
\end{eqnarray}
The latter relations are considered at each time, $t$, and express the trajectory of the charge through the guiding center coordinates $(t,X,\gamma,\mu,\varepsilon)$. Comparing such relations with the parametrized trajectories in (\ref{transf}), it is straightforward to identify $\alpha$ with $\alpha=(\gamma,\mu,\varepsilon)$. Specifically $\alpha_1=\gamma,\alpha_2=\mu$ and $\alpha_3=\varepsilon$. In (\ref{gctransf}), $\rho$ is the \emph{gyro-radius}  and $X$ is the guiding center position. The guiding center velocity is $\dot X=V$, if $V$ is computed at $X$.\\
The contra-variant guiding center coordinates are $Z^A=(t,X,\gamma,\mu,\varepsilon)$, with the index $A$ from $0$ to $6$. The coordinates $Z^A$ are contra-variant because they transform in the following way: if $Z^A=Z^A(z^B)$ then 
  \begin{equation}
  d Z^A=\frac{\partial Z^A}{\partial z^B} d z^B.
  \end{equation}
  In the guiding center description of motion the single particle lagrangian doesn't depend on $\gamma$, therefore,  a reduction of the complexity of the problem is achieved:  $\gamma$ is  cyclic and the conjugate variable, that is  the magnetic moment, $\mu$, becomes an invariant of motion. The lagrangian in (\ref{L0}) is rewritten by separating the guiding center part:
    \begin{eqnarray*}
 &&L_\mathrm{nr}(t,X+\rho,V+\sigma)=\\
 &&=\frac{(V+\sigma)^2}{2}+(e/m)(V+\sigma)\cdot (A+\delta_\rho A)-(e/m)(\Phi+\delta_\rho \Phi), 
  \end{eqnarray*}
  where $A(t,x)=A(t,X)+\delta_\rho A$ and $\Phi(t,x)=\Phi(t,X)+\delta_\rho \Phi$. Here, $V$ is always solution of equation (\ref{ohmslawV}) for ensuring that if $\mu\to 0$ then $V$ describes the velocity of a particle. The lagrangian is written in such a way that it is divided in what surely doesn't depend on $\gamma$, firsts three terms below, from what it should:
  \begin{eqnarray}
  \label{Lnr0}
  \nonumber
 &&L_\mathrm{nr}=\frac{V^2}{2}+(e/m)V\cdot A-(e/m)\Phi+ \\
 \nonumber
 && + P \cdot \sigma +(e/m) (V \cdot  \delta_\rho A- \delta_\rho \Phi)+\\ 
 &&+ \frac{\sigma^2}{2}+(e/m) (\sigma \cdot  \delta_\rho A).
  \end{eqnarray}
  It is  advantageous denoting $L_\mathrm{nr0}=V^2/2+(e/m)V\cdot A-(e/m)\Phi$, as the leading order lagrangian, with $L_\mathrm{nr1}=P \cdot \sigma +(e/m) [\rho \cdot (\nabla A)\cdot V- \rho \cdot \nabla \Phi]$, as the first order lagrangian, and the rest with $L_\mathrm{nr2}=L_\mathrm{nr}-L_\mathrm{nr0}-L_\mathrm{nr1}$. The reason is that the equation of motions are obtained only if the first order lagrangian vanishes. This it can be easily seen if $\rho$ and $\sigma$ are considered small. In such case $L_\mathrm{nr2}$ can be ignored and $L_\mathrm{nr1}=\rho\cdot(e/m)  [(\nabla A)\cdot V- \nabla \Phi-\dot{P}]+d(\rho\cdot P)/dt$. The equation of motion are obtained when it is required that $L_\mathrm{nr1}-(d/dt) (\rho \cdot P)=0$, if the identity $(\nabla A)\cdot V=V\times \nabla \times A+V\cdot \nabla A$ is used. This is equivalent to ask for the action to be at an extremal since, neglecting the small variation of the trajectory $\rho$, $L_\mathrm{nr}=L_\mathrm{nr0}$.
  \subsection{The guiding center gauge function}
 The single particle lagrangian is gauge independent and, for a gauge function $g$, then the transformation $A\rightarrow A- \nabla g$ and $\Phi \rightarrow \Phi +\partial_t g$ leaves the trajectories of motion unaltered. Indeed, together with the gauge transformation of e.m. potentials the lagrangian is shifted, $L_\mathrm{nr} \rightarrow L_\mathrm{nr}-(e/m)\dot g$, and the addition of a total time derivative doesn't affect the EL equations of motion. However, it is common practice to use a gauge function that depends on space and time only. Here, we use a gauge function that depends on the guiding center coordinates. It is worth noticing that quantum mechanics forbids the dependency of e.m. potentials from the velocities by limiting the domain of the gauge function only to the whole space-time. Even though the redefinitions of e.m. potentials doesn't affect the e.m. field (E and B), it is now possible to cancel or add some terms in the lagrangian that depend on all the variables of the whole phase space, like $\gamma$. The dependency of $\gamma$ in the single particle lagrangian can be \emph{manipulated} through an efficient choice of the gauge \cite{scott17}. Let's try to express the single particle lagrangian in the guiding center coordinates, for $A, B = 0, 1, 2, 3, 4, 6$: $\tilde{L}_\mathrm{nr}(Z^A,\dot{Z}^B)=\tilde{L}_\mathrm{nr}(t,X,\gamma,\mu,\varepsilon,V,\dot{\gamma},\dot{\varepsilon})$.  The simplest way to express the lagrangian in (\ref{Lnr0}) in the guiding center coordinates, $L_\mathrm{nr}=\tilde{L}_\mathrm{nr}(Z^A,\dot{Z}^B)$, is setting $g=g(Z^A,\dot{Z}^B)$ and asking for the following relation
  \begin{eqnarray}
  \nonumber
 &&(e/m)\dot g=- P \cdot \sigma -(e/m) (V \cdot  \delta_\rho A- \delta_\rho \Phi) +\\
 && - \frac{\sigma^2}{2}-(e/m) (\sigma \cdot  \delta_\rho A).
  \end{eqnarray} 
  Moreover, if $\partial_\gamma \dot g=0$ then the lagrangian
  \begin{equation}
  \tilde{L}_\mathrm{nr}= \frac{V^2}{2}+(e/m)V\cdot A-(e/m)\Phi -(e/m)\dot g
  \end{equation}
  doesn't anymore depend on $\gamma$.\\
  However, the former relation denotes something really important, the equivalence of 
  \begin{equation}
  \label{genefunc}
  p\cdot v-\varepsilon=P\cdot V-\mathcal{E}-(e/m)\dot{g},
  \end{equation}
  being $\varepsilon=v^2/2+(e/m)\Phi(t,x)$ and $\mathcal{E}=V^2/2+(e/m)\Phi(t,X)$.
 From the former relation it is clear the reason for indicating $(e/m) g$ as the \emph{(Lie) generating function for the guiding center transformation}. 
  In the non perturbative guiding center transformation it is chosen to set $\varepsilon=\mathcal{E}+(e/m)\dot{g}$ and the product $p\cdot v$  is conserved: $P\cdot V=p\cdot v$. From the linear dependency of the energy on $\dot g$, it seems that the energy of the particle should depends also on the chosen gauge. This is what it commonly happens because the energy is linearly dependent on the electric potential which itself is gauge dependent. However, it could be nice if the energy becomes independent from the gauge function, $g$. This means that $\partial_g \dot{g}=0$. In fact, we have just seen that for asking $\gamma$ to be cyclic in the lagrangian, then $\partial_\gamma \dot{g}=0$. Thus, a simple identification occurs: $g\propto \gamma $. Let's introduce the magnetic moment as the constant useful for identifying the gauge function with the gyro-phase:
  \begin{equation}
  \label{gfunc}
  g=(m/e)^2 \mu \gamma,
  \end{equation}
  then 
  \begin{equation}
  \label{fotoel}
  \varepsilon=\mathcal{E}+(m/e)\mu \dot{\gamma}.
  \end{equation}
  At the same time, it is found from the transformation in (\ref{gctransf}) that
  \begin{equation}
  \varepsilon=\mathcal{E}+\frac{\sigma^2}{2}+\sigma\cdot V+(e/m)\delta_\rho \Phi.
  \end{equation}
  Comparing the latter relations, the definition of the \emph{cyclotron (angular) frequency}, $\omega_c=\dot{\gamma}$, is:
  \begin{equation}
  \label{omegac}
  \omega_c=\frac{\sigma^2/2+\sigma\cdot V+(e/m)\delta_\rho \Phi}{(m/e)\mu},
  \end{equation}
  and it doesn't depend on $\gamma$, because $\gamma$ has been imposed to be cyclic for construction.
  It is quite easy to demonstrate that the magnetic moment is the constant of motion associated to the cyclic variable $\gamma$.
  In this way we have constructed a constant magnetic moment. Its constancy has not been explicitly derived but required for consistency from the following property: the gauge independency of electrodynamics.     \\The gauge transformation is 
  \begin{equation}
  A\rightarrow A -(m/e)\mu \nabla \gamma,\qquad \mbox{  and }\qquad \Phi \rightarrow \Phi+(m/e)\mu\partial_t \gamma
  \end{equation}
  also for the relativistic case.
 \subsection{The relativistic guiding center transformation}
 The relativistic  transformation is implicitly written as
  \begin{eqnarray}
\label{rgctransf}
x=X+\rho(t,X,\gamma;\mu,\varepsilon)\\
\nonumber
u=U(t,X;\mu,\varepsilon)+\nu(t,X,\gamma;\mu,\varepsilon),
\end{eqnarray}
being $u=x^\prime$, $U=X^\prime$ and $\nu=\rho^\prime$. 
The relativistic lagrangian is:
 \begin{equation}
L=-\frac{u^\alpha u_\alpha}{2}-\frac{e}{m} u^ \alpha A_\alpha(t,x)-\frac{1}{2}.
\end{equation}
Once $u^0=\gamma_v$ has been considered, it is convenient to introduce $U^0$ and $\nu^0$. The guiding center velocity is not the velocity of a charge, so that $U^\alpha U_\alpha \neq 1$ if $\nu^\alpha \neq 0$. Indeed, it is possible to ask for $\nu^\alpha \nu_\alpha=0$. In such case $\nu^\alpha$ is a light-like four-vector, as for a photon. Such correspondence is stressed writing $U^\alpha=w^\alpha-\eta^\alpha$, with $\eta^\alpha \eta_\alpha=0$, as for $\nu$, and $w^\alpha w_\alpha=1$, as for a charge. Now, it is possible to re-write the guiding center four-velocity transformation, $u^\alpha=U^\alpha+\nu^\alpha$, in the following way:
\begin{equation}
u^\alpha+\eta^\alpha=w^\alpha+\nu^\alpha.
\end{equation} 
The advantage is that it is possible to obtain the correspondence with the conservation of energy and momentum in  the \emph{Compton}-like scattering ($e^-+\gamma \rightarrow e^-+\gamma$).
It is worth noticing that the role of the guiding center is equivalent to the \emph{virtual particle} in particle physics, now. From such correspondence it  is possible to easily obtain the relation between $\nu^0$ and $\eta^0$ like for the frequencies involved in the \emph{Compton} scattering: if $\eta \cdot \nu=\eta^0\nu^0\cos \theta$ and $w^\alpha=(1,0,0,0)$, then
\begin{equation}
\eta^0=\frac{\nu^0}{1+\nu^0(1-\cos \theta)},
\end{equation} 
and
\begin{equation}
U^0=w^0-\eta^0=\frac{1-\nu^0 \cos \theta}{1+\nu^0(1-\cos \theta)},
\end{equation}
being $\nu_0=\pm|\nu|$ and, from $U^0+\nu^0=\gamma_v$,
\begin{equation}
\cos \theta= 1-\frac{\gamma_v-1}{\nu^0(\gamma_v-\nu^0)}.
\end{equation}
\\
 The gauge function, $g$, should be chosen chosen in such a way that the equivalence in (\ref{genefunc}) is replaced by
\begin{equation}
L=- p_\alpha u^\alpha=- P_\alpha U^\alpha- (e/m)g^\prime.
\end{equation}
Thus,  after a gauge transformation, the product  $u^\alpha p_\alpha$ is not an invariant anymore. Here, it is possible to anticipate what it will be crucial in the KK model: if you assign the values  $z^0=t$, $z=X$, $z^4=\gamma$ and $w_0=P_0=U_0+(e/m)\Phi$, $w=P=U+(e/m)A$ and $w_4=(m/e)\mu$, then 
\begin{equation}
L=- w_a z^{\prime a}, \qquad \mbox{for a=0,1,2,3,4.}
\end{equation} 
which is a scalar product in a space-time of five dimensions. Moreover, if you require that $w_5=w_6=0$ and $z^5=\mu$, $z^6=\varepsilon$, then you can also write
 \begin{equation}
L=- w_A z^{\prime A}, \qquad \mbox{for A=0,1,2,3,4,6.}
\end{equation} 
The latter is what is called the phase-space lagrangian from which it is possible to find the \emph{Hamilton's equations}. It is better to denote with a \emph{tilde} the guiding center phase-space lagrangian: $L=\tilde{L}(z^A,z^{\prime B})$, for $A, B = 0, 1, 2, 3, 4, 6$. As said in \cite{cary1983}, the reason for the vanishing of $w_5$ and $w_6$ is due to the fact  that $\varepsilon$ is the conjugate coordinate of $t$ and $(m/e)\mu$ is the conjugate coordinate of $\gamma$.\\
Now, the lagrangian is invariant at a glance with respect to general non-canonical phase-space coordinates transformations, that include also the gauge transformations.
\\
\section{Non-canonical lagrangian for the guiding center description}
Following the work of Cary and Littlejohn \cite{cary1983}, it is possible to find a \emph{lagrangian} derivation of the former guiding center description. The point here is to describe the hamiltonian mechanics using non-canonical variables on the extended phase space (position, velocity and time, hereafter). We start with  simple static case with time independent  fields. The (\emph{Maupertius}) principle of least action states that:
\begin{equation}
\delta W=\delta \int_{s_\mathrm{in}}^{s_\mathrm{out}} ds\, p(x) \cdot \frac{dx}{ds}=0.
\end{equation}
with $p(x)=u(x)+(e/m) A(x)$ and $p(x) \cdot \delta x=0$ at the end points. The EL equations are
\begin{equation}
\frac{dx}{ds} \times \nabla \times p(x) =0,
\end{equation}
which means that the velocity $u$ is parallel to $\nabla \times p(x)$ or
\begin{equation}
u=\lambda \nabla \times p(x),
\end{equation}
 re-obtaining the fundamental equation (\ref{claudio1}). \\
 The lagrangian $p(x) \cdot u$ is missing something. Now, we explicitly consider the time and the variation of the time dependent action:
 \begin{equation}
 \delta S=\delta \int_{s_\mathrm{in}}^{s_\mathrm{out}} ds\, \left[ p(t,x) \cdot \frac{dx}{ds}- \varepsilon(t,x) \frac{dt}{ds}\right]=0.
 \end{equation}
 The variation can be computed as
  \begin{eqnarray}
  \nonumber
 &&\delta S= \int_{s_\mathrm{in}}^{s_\mathrm{out}} ds\, \delta x \cdot \nabla \left[ p(t,x) \cdot \frac{dx}{ds}- \varepsilon(t,x)  \frac{dt}{ds}\right]+\\
 \nonumber
 &&+\int_{s_\mathrm{in}}^{s_\mathrm{out}} ds\, \delta t \partial_t  \left[ p(t,x) \cdot \frac{dx}{ds}- \varepsilon(t,x)  \frac{dt}{ds}\right]+\\
 \nonumber
 &&+\int_{s_\mathrm{in}}^{s_\mathrm{out}} ds\, \left[\delta \frac{dx}{ds} \cdot p(t,x)-\delta \frac{dt}{ds} \varepsilon(t,x)\right].
 \end{eqnarray}
 In the present notation, $\delta x= \rho$ and $\delta t= \rho^0$.
 In covariant notation, such variation is
   \begin{eqnarray}
  \nonumber
 &&\delta S= \int_{s_\mathrm{in}}^{s_\mathrm{out}} ds\, \rho^\alpha  u^\beta ( \partial_\beta p_\alpha -\partial_\alpha p_\beta)+\\
 \nonumber
 &&+(\rho^\alpha p_\alpha)|_{s_\mathrm{in}}^{s_\mathrm{out}}.
  \end{eqnarray}
  The extremals of the action, $\delta S=0$, for all the trajectories with $\rho^\alpha p_\alpha=0$ (at least at $s_\mathrm{in}$ and $s_\mathrm{out}$), are found if
 \begin{equation}
 \label{idlohmbis}
u^\beta ( \partial_\beta p_\alpha -\partial_\alpha p_\beta)=0,
 \end{equation}
 which is (\ref{idlohm}).
 \\
  Up to now, we have referred to the guiding particle as the particle satisfying (\ref{idlohm}), or (\ref{idlohmbis}), with null magnetic moment (and minimally coupled with the magnetic field). The same equation (\ref{idlohm}), or (\ref{idlohmbis}), is considered as the equation describing the \emph{guiding center velocity} if the particle has a non vanishing magnetic moment. In such case we use capital letters \emph{e.g.} for describing the position $X$, the velocity $V=\dot X$ and the four-momentum $P_\alpha=(\mathcal{E},-P)$, of the guiding center.\\
  We can add or subtract to the lagrangian a total world line derivative without changing the equation of motion and preserving the scalar value of the Lagrangian. We subtract  to $L=P\cdot U-\mathcal{E}U^0$ the total derivative of the following \emph{gauge function}:
\begin{equation}
\label{mygauge}
 g=(m/e)^2 \mu  \gamma,
 \end{equation}
 being $\mu$ the constant magnetic moment and $\gamma$ the gyro-phase.\\
 The new lagrangian is $\tilde L=P\cdot U-\mathcal{E}U^0- (e/m) g^\prime=P\cdot U-\mathcal{E}U^0-(m/e) \mu  \gamma^\prime $.
 With respect to the lagrangian $L= P\cdot U-\mathcal{E}U^0$, $\tilde L$   is known as (minus) the \emph{Routhian}, which is defined  through the \emph{Legendre} transformation of $L$ with respect to the cyclic coordinate $\gamma$:
 \begin{equation}
 -\tilde L \equiv (m/e)\mu \gamma^\prime- L.
 \end{equation} 
 The properties of $\tilde L$ are to combine the EL and the Hamilton's equations together for describing the motion:
  \begin{equation}
 \frac{d}{ds}\nabla_U \tilde L-\nabla_X \tilde L=0.
 \end{equation}
 and
 \begin{equation}
 \label{routheq}
 (e/m)\partial_\mu \tilde L=\gamma^\prime, \qquad \partial_\gamma \tilde L=-(m/e)\mu^\prime,
 \end{equation}
 respectively.
  The use of the \emph{Legendre} transformation for the cyclic variable has been used for describing the motion with the coordinates $\gamma,\mu$ instead of $\gamma,\gamma^\prime$. Being $\gamma$ cyclic, $\mu^\prime=0$ in (\ref{routheq}).\\
\subsection{Non canonical Hamilton's equations of motion}
The present and the next paragraphs are quite technical, but it is important to describe what concerns the dimensional reduction of a system. Historically, the dimensional reduction was a technique used to attack a complicated problem by progressively reducing it in order to reach a resolvable system.  In gyro-kinetic the dynamic of the particle is separated from the fast gyro-motion reducing the analysis to the dynamic of the guiding center (if fluctuations are turned off). In the KK mechanism \cite{kk1,kk2}, the same particle dynamic, now extended to consider also the presence of a gravitational field, is reduced from a five-dimensional to a four-dimensional space-time, leaving the $5^{th}$ dimension unobservable. Thus, the \emph{Routhian} reduction scheme \cite{routh} is a method implemented to describe a mechanical system where the reduction is made to suppress an angle coordinate after a smart change of variables. We will see how all these reduction schemes can be seen as different approaches for disregarding the gyro-phase from the equations of motion. However, in the present section we want to show why it is possible to reduce the dimensionality of a system by cutting out a coordinate from the description of motion.\\
The idea, originally proposed by \cite{cary1983} even if applied only at the perturbative approach, was to properly use non canonical coordinates in Hamiltonian mechanics for simplifying a problem. Starting from requiring that the Lagrangian is a scalar, it is written as the scalar product between coordinates and momenta. The coordinates for describing the motion can be changed together with the conjugate momenta  but by taking care that such transformation must not change the scalar value of the Lagrangian, which means that a relativity principle holds. As for example,  
 the 1-form associated to the guiding center \emph{lagrangian} is 
\begin{equation}
\label{tildeLform}
\tilde L d \tilde s= P dX -(m/e)\mu d\gamma-\mathcal{E} dt.
\end{equation}
For such system the motion is described by the variables $z^a=(t,X,\gamma)$, with index $a$ from $0$ to $4$, so that the world line coordinate, $\tilde{s}$, is function of $z^a$: $\tilde{s}=\tilde{s}(z^a)$. Moreover, the conjugate momenta, $w_a$, are easily introduced consistently with the lagrangian in (\ref{tildeLform}): $w_a=(\mathcal{E},-P,(m/e)\mu)$. Now, 
\begin{equation}
\label{KKform}
\tilde L =-w_a z^{\prime a}, \qquad \mbox{for}\qquad a=0,1,2,3,4.
\end{equation}
  However, following the analysis done in  \cite{cary1983}, it is more convenient to extend  the description of motion to the whole extended phase space.  The reason is that for charge motion the most useful coordinates appear like a mixture between positions and velocities, as for the canonical four-momentum. It is useful to consider the lagrangian in ({\ref{tildeLform}}) as the reduced lagrangian of the entire lagrangian that operates on the extended phase space, $\hat L$. Now the indexes, A,B, ..., go from $0$ to $6$ and the generalized coordinate is $z^A=(t,X,\gamma,\varepsilon, \mu)$ which includes also the independent coordinates $\varepsilon$ and $\mu$. It is worth noticing that we are adopting non-canonical coordinates. Here, we will refer to  $z^A=(t,X,\gamma,\varepsilon, \mu)$ as the guiding center coordinates. 
As before, it is possible to associate a set of conjugate momenta to such variables. The new co-momenta are $w_A=(\mathcal{E},-P,(m/e)\mu,0,0)$, as similarly chosen in \cite{cary1983} for a different problem. However, it is worth noticing that $w_A=w_A(z^B)$ is function of the non-canonical coordinates so that $\hat L= \hat L(z^A)$. Thus, the lagrangian can be written as 
\begin{equation}
\label{PCform0}
\hat L d \hat s= -w_A dz^A, \qquad \mbox{for}\qquad A=0,1,2,3,4,5,6.
\end{equation}
 The scalar character of the lagrangians, (\ref{KKform}) like (\ref{PCform0}), is always preserved and it is possible to change coordinates from $z^A \to Z^A$ and $w_A \to W_A$ leaving unaltered the \emph{Poincar{\'e}-Cartan form}: $-w_A(z^B) dz^A=-W_A(Z^B) dZ^A$. This means that the \emph{principle of relativity} is generalized to the extended phase space: a change of coordinates of the extended phase space preserves the physics.\\
The EL equations for (\ref{PCform0}) are 
\begin{equation}
\label{generalohm}
\omega_{AB} \frac{dz^B}{d\hat s}=0,
\end{equation}
with
\begin{equation}
\label{tensorfreq}
\omega_{AB}=\partial_Aw_B-\partial_B w_A.
\end{equation}
Multiplying eq. (\ref{generalohm}) for $d\hat s/dt$, it is found  what it can be called the \emph{velocity law} (compare (\ref{idlohm}) with  (\ref{generalohm})) in 7 dimensions (or 5 dimensions if the motion is described through $\varepsilon$ and $\mu$, if $\varepsilon^\prime=\mu^\prime=0$). It is worth noticing that a \emph{canonical Maxwell tensor} in 7 dimensions is proportional to $\omega_{AB}$. The generalized angular frequency tensor, $\omega_{AB}$, is known as the \emph{Lagrange} tensor. The Lagrange tensor is expressed by  the Lagrange's brackets:
\begin{eqnarray}
 &&[z^A, z^B]\equiv \partial_A w_C \partial_B z^C -\partial_B w_C \partial_A z^C=\\
 \nonumber
&&=\partial_A w_C \delta^C_B -\partial_B w_C \delta^C_A=\omega_{AB} 
 \end{eqnarray}
where $[z^A, z^B]$ are the \emph{Lagrange brackets}.\\ 
It is convenient to normalize $\dot z^0=1$ in (\ref{generalohm}), which means choosing $z^0=t$. Three properties of  motion must be reminded: 1) $\mathrm{det}\, (\omega_{AB})=0$, from (\ref{generalohm}), 2) the gauge invariance of motion if $w_A\to w_A+\partial_A g$, from (\ref{tensorfreq}), and 3) in (\ref{generalohm}) the case $A=0$ is redundant due to the antisymmetry of $\omega_{AB}$.\\  Equation (\ref{generalohm}) can be arranged to
\begin{equation}
\label{hamilton0}
\omega_{A0}+\sum_{B=1}^6\omega_{AB} \dot z^B=0,
\end{equation}
with 
\begin{equation}
\label{omegaA0}
\omega_{A0}=\partial_A w_0-\partial_t w_A=\partial_A \mathcal{E}-\partial_t w_A.
\end{equation}
For obtaining the Hamilton's equations of motion it occurs introducing the antisymmetric \emph{Poisson} tensor, $J^{AB}$, with the property that
\begin{equation}
\sum_{C=1}^6J^{AC}\omega_{CB}=-\delta^A_B, \mbox{if }A\neq 0 \, \mbox{and} \, B \neq 0
\end{equation}
 Now, the expression in (\ref{hamilton0}), with equation  (\ref{omegaA0}), becomes
\begin{equation}
\label{hamilton1}
\dot z^A=\sum_{C=1}^6J^{AC}\left( \partial_C \mathcal{E}-\partial_t w_C \right) \qquad \mbox{ if } \,A\neq 0.
\end{equation}
The latter are the Hamilton's equations of motion for non-canonical coordinates and it can be reduced to
\begin{equation}
\label{hamilton2}
\dot z^A=\{z^A,\mathcal{E}\}+\partial_t z^A.
\end{equation}
if canonical coordinates are employed, being $\{z^A,z^B\}$ the \emph{Poisson brackets}.

Within the guiding center description, when guiding center coordinates, $z^A=(t,X,\gamma,\mu,\varepsilon)$ and $w_A=(\mathcal{E},-P,(m/e)\mu,0,0)$, are used,
the Lagrange tensor is:
\begin{equation}
\begin{vmatrix} 
  (e/m)F_{\alpha \beta}c & 0 & -\partial_\mu P_\alpha & -\partial_\varepsilon P_\alpha\\
0 & 0 & -m/e & 0\\
\partial_\mu P_\beta & m/e & 0 & 0\\
\partial_\varepsilon P_\beta & 0 & 0 & 0\\
\end{vmatrix}
\end{equation}
The equations of motion, from (\ref{generalohm}), are
\begin{equation}
\begin{vmatrix} 
  (e/m)F_{c\alpha \beta} & 0 & -\partial_\mu P_\alpha & -\partial_\varepsilon P_\alpha\\
0 & 0 & -m/e & 0\\
\partial_\mu P_\beta & m/e & 0 & 0\\
\partial_\varepsilon P_\beta & 0 & 0 & 0 \\
\end{vmatrix}
\begin{vmatrix}
V^\beta \\ \dot \gamma \\ \dot \mu \\ \dot \varepsilon \\
\end{vmatrix}=0
\end{equation}
Explicitly, the system of equations of motion is:
\begin{eqnarray}
&&(e/m)F_{c\alpha \beta} V^\beta -\dot \mu \partial_\mu P_\alpha  -\dot \varepsilon \partial_\varepsilon P_\alpha=0 \\
&&-(m/e)\dot \mu=0\\
&&V^\beta \partial_\mu P_\beta +(m/e)\dot \gamma=0\\
\label{redund}
&&V^\beta \partial_\varepsilon P_\beta=0,
\end{eqnarray}
where the equation (\ref{redund}) is redundant for the antisymmetry of $F_{c\alpha \beta}$. If also $\dot \varepsilon=0$ then
\begin{eqnarray}
&&E_c+V\times B_c =0 \\
&&\dot \mu=0\\
&&V \cdot \partial_\mu P =(m/e)\dot \gamma(1-V^0)\\
&&V\cdot \partial_\varepsilon P=V^0,
\end{eqnarray}
If $V^0=1$ then $V \cdot \partial_\mu P =0$ and $V\cdot \partial_\varepsilon P=1$.\\
It is worth noticing that all the lagrange brackets involving $\gamma$ and $\mu$ are null,  $[\mu,\gamma]$ apart, which is equal to $m/e$. This is the reason that allows reducing the particle motion ignoring the \emph{gyro-phase} coordinate, $\gamma$, which is said \emph{cyclic}. \\
\subsection{Leading order non-relativistic  guiding center transformation}
In the perturbative approach of the guiding center transformation a completely different procedure is often used. Moreover, the perturbative approximation is treated with the highly technical Lie-transformation method without solving some ambiguities.  For such reason it is often hard to overcome the leading order approximation. However, in the present work, we do not consider a comparison between the two distinct methods, the perturbative and the non-perturbative one. Here, we need only the first order approximation for explaining why $\mu$ is the magnetic moment and $\gamma$ is the gyrophase. \\
The guiding center lagrangian used in the perturbative approach is the leading order approximation of $L_{gc}$, which is the lagrangian in (\ref{PClagr}) associated to the \emph{Poincar{\'e}- Cartan} one-form:
\begin{equation}
L_{gc}=L_\mathrm{nr}+(e/m)\dot{g}=P\cdot \dot{X}-\mathcal{E}.
\end{equation}
Explicitly,
\begin{equation}
L_{gc}=p\cdot \dot{x}-\varepsilon+(m/e)\mu \dot{\gamma},
\end{equation}
being $g=(m/e)^2\mu \gamma$, the guiding center gauge function in (\ref{mygauge}).
Setting $\varepsilon=\mathcal{E}+(m/e)\mu \dot{\gamma}$, as in (\ref{fotoel}), then 
\begin{equation}
\label{lgc}
L_{gc}=P\cdot \dot{X}-\varepsilon+(m/e)\mu \dot{\gamma}.
\end{equation}
The orderings, which are commonly employed, are the ones that allow to consider the particle close to the magnetic field line, in such way that field lines deviate only linearly from being straight and uniform (this is quite a rough approximation but almost always used). Within such orderings,  the charges are  gyrating circularly around the guiding center. Once the tern of unit vectors, $e_1\cdot e_2 \times b_{(0)}=1$, are defined with $b_{(0)}\cdot B=|B|$ parallel to the magnetic field,  the guiding center is considered mostly moving in the parallel direction of the magnetic field in such a way that $P$ is substituted with $P\approx v_\| b_{(0)}+(e/m) A(t,X)$.   The gyro radius can be written as
\begin{equation}
\rho\approx a_{(0)}=\rho_L(e_1 \cos \gamma -e_2 \sin \gamma),
\end{equation}
 with the constant \emph{Larmor radius}, $\rho_L$.
 It is worth noticing that $\gamma$ is the angle in the cylindrical representation of the velocity space. If $\dot \gamma=(e/m)|B|$, which is the important \emph{cyclotron frequency}, then
\begin{equation}
v_\perp \approx \dot{a}_{(0)}=-\rho_L\dot \gamma (e_1 \sin \gamma +e_2 \cos \gamma)=(e/m)a_{(0)}\times B.
\end{equation}
Commonly the electric potential is neglected and the energy of the charge is the only kinetic energy:
\begin{equation}
\varepsilon=v^2/2.
\end{equation}
Moreover, the problem is often considered static: $A=A(X)$ with $\partial_t A=0$.
The single particle lagrangian,
\begin{equation}
L=[v_\| b_{(0)}+v_\perp+(e/m)A]\cdot \dot x-\varepsilon,
\end{equation}
is approximated by
\begin{eqnarray}
&& L\approx   (v_\| b_{(0)}+\dot a_{(0)})\cdot (\dot X+\dot a_{(0)})+\\
\nonumber
&&+[(e/m)A+(e/m)a_{(0)} \cdot \nabla A]\cdot (\dot X+\dot a_{(0)})-\varepsilon,
\end{eqnarray} 
which is regrouped and simplified to
\begin{eqnarray}
&& L\approx   [v_\| b_{(0)}+(e/m)A]\cdot \dot X-\varepsilon\\
\nonumber
&&+\dot{a}_{(0)}\cdot [(e/m)A+\dot{X}]+(e/m)a_{(0)} \cdot (\nabla A)\cdot \dot X\\
\nonumber
&&+(e/m)a_{(0)} \cdot (\nabla A)\cdot \dot a_{(0)}+\dot{a}^2_{(0)}.
\end{eqnarray} 
The terms in the first row are independent on $\gamma$, the ones in the second row depend on $\gamma$ and are not negligible, the ones in the third row also depend on $\gamma$ but they are very little. Now,  it is possible \cite{littlejohn83} to find a gauge function, $\mathcal{S}=-a_{(0)} \cdot(e/m)A-(e/m)a_{(0)} \cdot (\nabla A)\cdot a_{(0)}/2 $, for expressing the lagrangian as
\begin{equation}
L\approx   [v_\| b_{(0)}+(e/m)A]\cdot \dot X+\dot{a}^2_{(0)}/2-\varepsilon+d\mathcal{S}/dt.
\end{equation} 
as shown in \cite{scott17}.\\
Finally, defining the magnetic moment as
\begin{equation}
\mu\approx \frac{v_\perp^2}{2|B|},
\end{equation}
then $L_{gc}=L-d\mathcal{S}/dt$ is 
\begin{equation}
L_{gc}\approx [v_\| b_{(0)}+(e/m)A]\cdot \dot X+(m/e)\mu \dot{\gamma}-\varepsilon,
\end{equation}
which is the lagrangian obtained in (\ref{lgc}) if $P\approx v_\| b_{(0)}+(e/m) A(t,X)$. At the same ordering, being the energy quadratic, the guiding center energy is $\mathcal{E}\approx v_\|^2/2$ in such a way that the total energy is $\varepsilon=\mathcal{E}+(m/e)\mu\dot{\gamma}\approx v_\|^2/2+\mu|B|.$
\subsection{General comments}
The guiding center coordinates in the presence of a magnetic field, similarly to the center of mass coordinates in a gravitational field, describe the origin of the reference frame where positions, velocities and time are efficiently measured, so that 
\begin{eqnarray*}
\label{gctransf1}
x=X+\rho(\gamma)\\
u=U+\nu(\gamma)\\
t=t_b+\tau(\gamma).
\end{eqnarray*}
 It is worth noticing that the latter equation is often written in plasma physics as $t=t_\mathrm{slow}+t_\mathrm{fast}(\gamma)$, so dividing what depends on slow variations  from what depends on fast variations. In the present analysis $t_b$ is considered as a reference time  which it can also be used for obtaining the Abraham-Lorentz-Dirac force \cite{lorentz,adl}:
  \begin{equation}
 \ddot X(t) \approx \ddot X(t_b)+\tau \dddot X(t_b).
 \end{equation}
 It is worth to note that the guiding center transformation is simply a translational transformation on the extended  phase space. All the coordinates are translated by a quantity depending  on $\gamma\in S^1$. This property allows the following new definition to emerge: \emph{the guiding center reference frame is the particular reference frame where the particle moves in a closed orbit with a periodic motion} \footnote{The same definition pertains also to the gyrocenter when e.m. fluctuations are taken into account, see section VII.}. The efficiency of describing the general motion is only because the orbit is reduced to a closed loop parametrized by the angle $\gamma$. In order to reach such reference system we must subtract the relativistic guiding center velocity $U$ from $u$ and also shift the position of the particle to the guiding center position $X$.   
 In the guiding center reference frame it is possible to observe that the particle is gyrating in a closed loop with the cyclotron frequency. \\
If the manifold of the extended phase-space is not flat, then the above translations must be considered as if the quantities depending on $\gamma$ are parallel transported over the manifold.
\section{Kaluza-Klein solution}
 The coordinates $z^A$ with $A=0,1,2,3,4,5,6$, introduced in the previous section, belong to the extended phase space. As for general relativity, where a geometry is given to the space-time, in this section a geometry is given  to the extended phase-space. \\

We have seen that in the presence of  e.m. fields, it is useful to describe the motion in guiding center coordinates, $z^A=(t,X,\gamma,\mu,\varepsilon)$. For accuracy, the guiding center transformation is the map, $\mathcal{T}$, that allows to describe particles through the guiding center coordinates, $\mathcal{T}: (t,x,p)\rightarrow (t,X,\gamma,\mu,\varepsilon) $. It is worth noticing that the vector $X$ indicates the position of the guiding center, not of  the particle. If $\mu\neq 0$ then the particle is elsewhere from $X$.\\
     The KK mechanism was used in the past to explain the presence of gravitation and electromagnetism thanks to the addition of, at least, a new coordinate of spacetime. The KK model can be obtained from a \emph{Hilbert-Einstein} (HE) action extended to a space-time of five dimensions as reminded in the appendix. However, in the present approach, we adopt the same mechanism, in which the new dimension is a coordinate that belongs to the velocity space. In fact, the $5^{th}$ dimension is identified with the gyro-phase coordinate, $\gamma$. As a consequence we are changing the paradigm of the general relativity theory that only takes into account the space-time geometry. Thus, if you want to describe gravity then you can only consider the geometry of space-time, whilst if you want to describe gravity plus electromagnetism  you have to consider the geometry of the extended phase space.   Mathematically, it is not so difficult to extend the general relativity formalism to five or more (seven) dimensions. However, the physical interpretation of an \emph{Einstein equation} in extended phase space, is quite unusual to be exposed in the present work. What is proposed here is a minimal change of the KK model and the use of the relativistic guiding center transformation. In this section we leave the \emph{Minkowski} metric for a pseudo-\emph{Riemannian} one.\\
     Let's start from the \emph{Poincar{\'e}-Cartan} one-form in (\ref{KKform}): $\hat{L} d \hat{s}= -w_A dz^A$, for $A=0,1,2,3,4,5,6$.  The same one-form can be written as
\begin{equation}
\label{PCform2}
\hat {L} d\hat{s}= -\hat{g}_{AB} w^B dz^A,
\end{equation} 
being $\hat L$ a scalar quantity and where $\hat{g}_{AB}$ is the metric tensor with the property that $w_A \equiv \hat{g}_{AB} w^B$. Thus,  $w^B$ are the \emph{contravariant} momenta.
  Once the metric tensor is appeared, it is possible to apply a variational principle for finding it. For this reason, we consider a \emph{lagrangian density over the extended phase space} where the single particle lagrangian is multiplied for the distribution of masses and, then, added to the HE lagrangian in extended dimensions. In the following lagrangian density, 
 \begin{equation}
\label{PCEH}
 \ell a = f_m \hat{L} -\frac{ \hat{\mathcal{R}}}{16 \pi \hat{G}},
\end{equation} 
$f_m$ is the scalar distribution function of masses, for simplicity only one species is considered; $\hat{G}$ and $\hat {\mathcal{R}}$  are the gravitational constant and the \emph{scalar curvature} for the extended phase space, respectively. The scalar curvature is defined as usual: 
\begin{equation}
\hat {\mathcal{R}} =\hat{g}_{AB} \hat{Ric}^{AB},
\end{equation}
again,  $\hat{Ric}^{AB}$ is the \emph{Ricci tensor} in the extended phase space which is furnished of a \emph{Levi-Civita connection}. The lagrangian, (\ref{PCEH}), is a \emph{lagrangian density} over the extended phase-space and the action is computed from the integration of  $\ell a$ over the extended phase space. If $\sqrt{|\hat{g}|}$ indicates the square root of minus the determinant of the extended phase space metric, then the extended phase space volume element, $d\mathcal{M}$, can be written as:
\begin{equation}
\label{Velem}
d\mathcal{M}=\sqrt{|\hat{g}|}\, d^7 z, 
\end{equation}
 if the guiding center coordinates are used then $d^7 z=dt \,d^3 X \,d\gamma \,d\varepsilon \,d\mu$.
Explicitly, the action is:
\begin{equation}
\label{7action}
S=\int \ell a \,d \mathcal{M},
\end{equation}
which is a definite integration in a domain $\partial\mathcal{M}$ of the extended phase space.
It is possible to separate in $\ell a$ the effects of different contributions. A \emph{matter lagrangian distribution}:
\begin{equation}
\ell a_m = -f_m,
\end{equation}
a  \emph{field lagrangian distribution}:
\begin{equation}
\ell a_\mathrm{f}=- \frac{\hat{\mathcal{R}}}{16\pi \hat{G}},
\end{equation}
and an \emph{interaction dynamics lagrangian distribution}: 
\begin{equation}
\ell a_\mathrm{id}= f_m (1+\hat L ), 
\end{equation}
The distribution of masses, $f_m$ is taken as a scalar function: $f_m=f_m(z^A)$\footnote{We are implicitly imposing that matter cannot be created nor destroyed.}.\\
 Within the  guiding center description, $f_m$ indicates the presence of a particle of mass $m$ with guiding center coordinates $(t,X,\gamma,\mu,\varepsilon)$. 
   
 The particle described by $f_m$ must be counted only once to obtain the total mass, $M$, of the system. The following equivalence chain of integrations is assumed for the \emph{matter action}, $S_m$: 
\begin{equation}
\label{lam}
S_m= -\int f_m \sqrt{|\hat{g}|} d^7 z=-\int \rho_m \sqrt{-g} dt d^3X=-\int M \,d \hat s,
\end{equation}  
where $\rho_m$ is the mass density and, above all, $\sqrt{-g}$ is the square root of minus the determinant of the space-time metric. In fact, if you call $J_\mathcal{P}$ the quantity $\sqrt{|\hat{g}|}/\sqrt{-g}$, then:
\begin{equation}
\label{dens}
 \rho_m= \int f_m J_\mathcal{P} \,d\gamma \,d\varepsilon \,d\mu.
\end{equation}
The density of masses is obtained from the integration of the distribution of masses in the velocity space. If you introduce unspecified velocities or momenta, $\mathcal{P}$, with the only property that allows to write the latter velocity space  volume element:
\begin{equation}
d^3\mathcal{P}=J_\mathcal{P} \,d\gamma \,d\varepsilon \,d\mu,
\end{equation}
then the former integral is written in the usual form:
\begin{equation}
 \rho_m= \int f_m d^3\mathcal{P}.
\end{equation}
Concerning the \emph{fields action}, $S_\mathrm{f}$, we wish to have:
\begin{eqnarray}
\label{laHE}
&&S_\mathrm{f}= -\int \frac{\hat{\mathcal{R}}}{16\pi \hat{G}} \sqrt{|\hat{g}|} d^7 z=\\
\nonumber
&&=-\int \frac{F_{\alpha \beta} F^{\alpha \beta}}{4} \sqrt{-g} dt d^3X - \int \frac{R}{16\pi G} \sqrt{-g} dt d^3X,
\end{eqnarray} 
 In order to obtain the latter result we will use the KK mechanism. However, before doing that, we are interested in studying the \emph{interaction dynamics action} $S_\mathrm{id}$, that should be expressed by:
 \begin{equation}
 \label{laI}
S_\mathrm{id}= \int f_m (1+\hat L )\sqrt{|\hat{g}|} d^7 z=-\int  A_\alpha J^\alpha  \sqrt{-g} dt d^3X,
\end{equation} 
where $J^\alpha$ is the charge  four-current density which is a field depending on $(t,X)$. The former equation will be obtained in the forthcoming subsection.
 It is worth noticing that, if the above equations for $\ell a_m$, $\ell a_\mathrm{f}$ and for $\ell a_\mathrm{id}$, defined in (\ref{lam}), (\ref{laHE}) and (\ref{laI}), respectively, are considered, once $\ell a$ is integrated in the velocity space,  then the following lagrangian density appears:
\begin{equation}
\label{Ldens}
\mathcal{L}=- \rho_m-A_\alpha J^\alpha- \frac{F_{\alpha \beta} F^{\alpha \beta}}{4}-  \frac{R}{16\pi G}.
\end{equation}
The latter is exactly the lagrangian density used for describing the presence of (e.m. interacting) matter as source of a gravitational field, which gives the \emph{Einstein} equation, together with a charge  four-current density as source of an e.m. field, which gives the \emph{Maxwell} equations. 
 \subsection{The \emph{misleading} symmetry}

In \emph{lagrangian mechanics} the symmetries of a system are expressed by the invariance of the lagrangian under the considered transformations. In relativity, the conservation of the energy-momentum tensor,  $T^{\alpha \beta}$, is fundamental. The conservation of $T^{\alpha \beta}$ is due to the symmetry of the lagrangian under the spacetime translation: $X^\alpha \to x^\alpha=X^\alpha+\rho^\alpha$. This is also true if we explicitly take, $X^\alpha=(t_b,X)$ and $\rho^\alpha=(\tau,\rho)$; so that, $x=X+\rho$ and $t=t_b+\tau$. If the manifold is not flat the translation is expressed by the parallel transport. \\
For our needs, the single particle lagrangian, $L=p\cdot u-\varepsilon \gamma_v$ can be written with a null magnetic moment term: $L=p \cdot u-\varepsilon \gamma_v+(m/e)\mu_0 \omega_{0}$, if $\mu_0=0$. Now, the guiding center transformation leaves unaltered the form of the lagrangian. In the non perturbative guiding center transformation,  the momentum of the particle, $p\rightarrow P$, becomes the guiding center momentum computed at the guiding center $X$ and at the time $t$, whereas the particle relativistic velocity, $u \rightarrow U$, becomes the relativistic guiding center velocity $U$. Moreover, the null magnetic moment $\mu_0 \rightarrow \mu$ becomes a positive magnetic moment so that the gyro-phase $\gamma$ becomes meaningful (because if $\mu=0$ then $\gamma$ is singular). The single particle lagrangian under such transformation, is
\begin{eqnarray}
L=p \cdot u-\varepsilon \gamma_v=p\cdot u-\varepsilon \gamma_v+(m/e)\mu_0 \omega_{0} \rightarrow \\
\nonumber
 \rightarrow \hat{L}=P\cdot U-\mathcal{E} U^0+(m/e)\mu \gamma^\prime,
\end{eqnarray}
which is the \emph{guiding center Lagrangian}, already seen in the former section. \\
In the relativistic approach we haven't yet considered a relation as $(m/e)\mu \dot\gamma=\varepsilon-\mathcal{E}$, used for defining the frequency, $\dot \gamma$. Such relation was used in the non relativistic case for obtaining $p\cdot v=P\cdot V$. In the relativistic case, another relation  is chosen that allows to write $L=-1+(e/m)u^\alpha A_\alpha(t,x)=\hat L$ with
\begin{equation}
\hat  L=-1+(e/m)U^\alpha A_\alpha (t,X),
\end{equation}
which is the same form of $L$. This means that $u^\alpha A_\alpha(t,x)=U^\alpha A_\alpha (t,X)$ is preserved. From $u^\alpha p_\alpha=U^\alpha P_\alpha+(m/e) \mu \gamma^\prime$, it was immediately found that the required condition is reached if
\begin{equation}
(m/e) \mu \gamma^\prime=1-U^\alpha U_\alpha.
\end{equation}
The latter relation is also more interesting if $(m/e)\mu \gamma^\prime=U^4U_4$, where $U^4=z^{4\prime}=\gamma^\prime$ and $U_4$ is firstly defined as $U_4\equiv w_4=(m/e)\mu$. In such way that
\begin{equation}
\label{misleading}
U^aU_a=1, \qquad \mbox{ for } a=0,1,2,3,4.
\end{equation}  
Moreover, if the relation $w_a=U_a+(e/m) A_a$ is used, then $A_4=0$ for consistency: there is not a $5^{th}$ component of the e.m. potential.
The symmetry that leaves invariant the form of $L=-1+(e/m)u^\alpha A_\alpha(t,x)=-1+(e/m)U^\alpha A_\alpha (t,X)$ is said misleading because there is no way, starting from the lagrangian from the dynamics),  to distinguish particle's coordinates from guiding center's coordinates. The only chance for appreciating the difference is by measuring the dispersion relation: from kinematics, the particle has $u^\alpha u_\alpha=1$ whilst the guiding center doesn't, $U^\alpha U_\alpha \neq1$. 
 If we suppose to observe a helicoidal trajectory made by the motion of a charged particle in a given e.m. field, then such trajectory could be considered a solution of motion. However, it is possible to zoom on the trajectory, by increasing the sensibility of detectors, and discover that the simple helicoidal trajectory is made by another sub-helicoidal motion, as shown in the cartoon of figure 1. At first sight the trajectory of the particle has been confused with the trajectory of the guiding center. 
 \begin{figure}[htbp]
 \begin{center}
 \mbox{
\includegraphics[width=8.5cm]{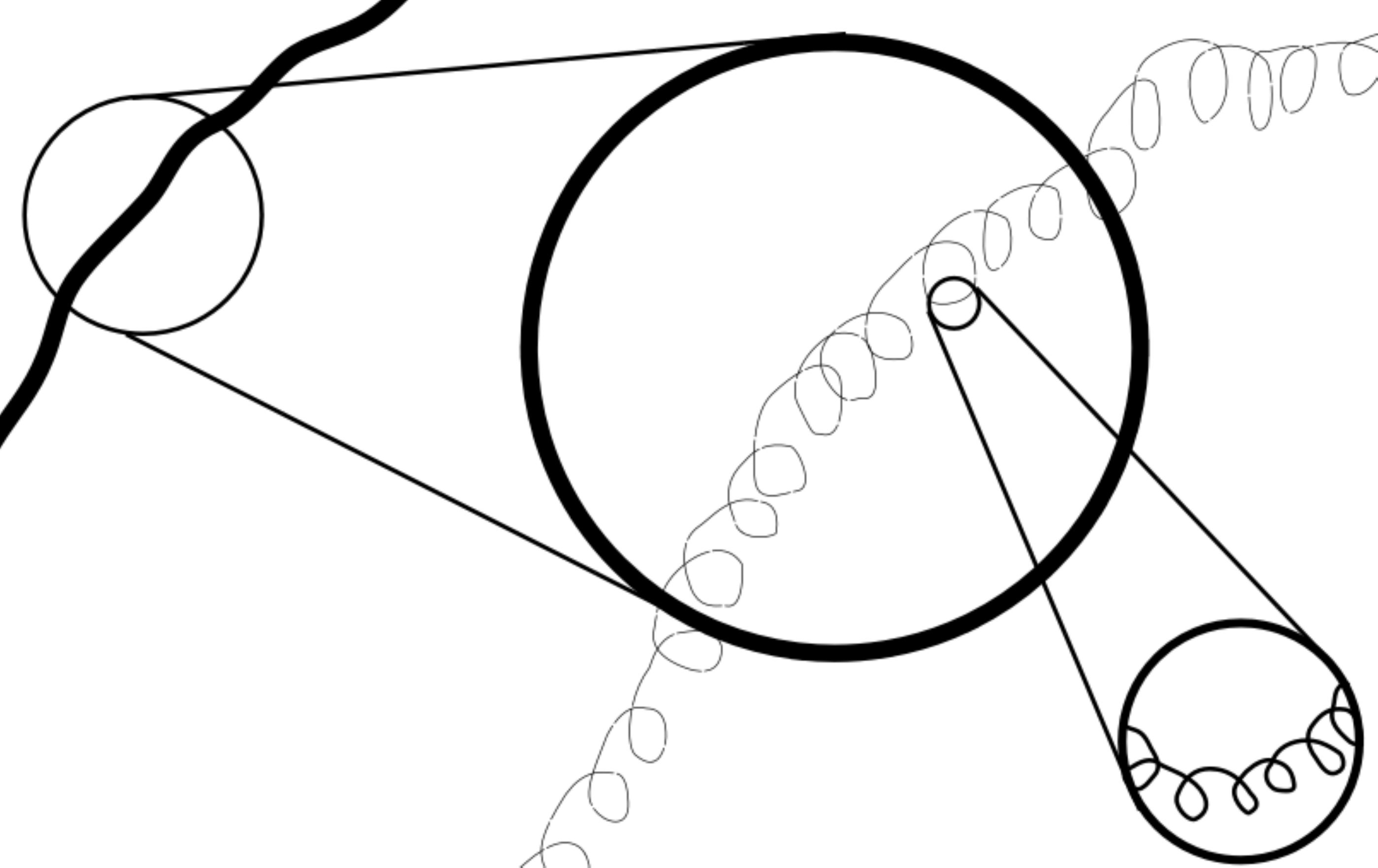}
}
\caption{The same trajectory in space of a charged particle in an e.m. field observed with three different resolutions. At first sight each curve can be understood as the particle's trajectory but it could also not be.}
\end{center}
\label{orbits}
\end{figure}
  Moreover, such misinterpretation can be iterated (with some constrains \emph{e.g.} the velocity cannot overcome the speed of light), so that the sub-helicoidal motion can, once again, hide another subsub-helicoidal motion at a finer scale. Similarly to a fractal, when the magnetic field differs from being constant and uniform a family of solutions enriches the extended phase space of helicoidal trajectories made by other helicoidal trajectories. It is worth noticing that realistic magnetic fields are never constant and uniform and, moreover, any realistic detector doesn't have infinite resolution.\\
   The approximation of considering the guiding center motion instead of the particle motion is said \emph{drift approximation} and, if applied with criteria, it becomes the zero-th order approximation in all the gyro-kinetic codes used for studying magnetic confined plasmas for controlled fusion through a kinetic perspective.\\
   It is worth noticing that there is another interpretation where many trajectories are described by the same motion of a representative guiding center. The latter interpretation is possible because we have considered all the trajectories with free initial and final conditions in the variational approach  in Section II. In fact, if we impose with (almost) certainty the initial and the final values of the particle's coordinates then there is only (almost)  a unique solution of motion, whilst for an initial and final uncertainty, there is the possibility to have many trajectories that differs by the value of the magnetic moment and by the initial value of the gyrophase.   Thus, the  question is: "what are the trajectories that minimize the action and are also well represented by the Lorentz's force law?", instead of being "what is the trajectory that  minimizes the action and is solution of the Lorentz's force law?". Those trajectories are indistinguishable and can be resolved only after a measurement, like for the collapse of a quantum state into a physical eigenstate in quantum mechanics. Indeed, the only way for distinguishing a guiding center from a particle is  from the misleading condition in (\ref{misleading}) that pertains to kinematics, being expressed by the \emph{Lorentz violation}, $U^\alpha U_\alpha \neq1$. The  dynamics is still preserved by the same lagrangian.\\     
 Thanks to the misleading condition, equation (\ref{misleading}), it is very easy to show that the action $S_\mathrm{id}$ takes the desired form (\ref{laI}) when the guiding center coordinates are used. In fact, $\hat L =-1+(e/m) A_\alpha U^\alpha$ and
 \begin{equation}
S_\mathrm{id}= \int f_m (1+\hat L )\sqrt{|\hat{g}|} d^7 z=-\frac{e}{m}\int  \rho_m A_\alpha \bar{U}^\alpha  \sqrt{-g} \,dt \,d^3X.
\end{equation} 
 If $J^\alpha=(e/m)\rho_m \bar{U}^\alpha$, being
 \begin{equation}
 \rho_m \bar{U}^\alpha =\int f_m U^\alpha d^3\mathcal{P},
 \end{equation}
  then the former is exactly the relation in (\ref{laI}).\\
 We have just seen that the guiding center transformation, which is a particular \emph{local} translation in the extended phase space, is a symmetry because it leaves the same lagrangian form. In analogy to what happens for the \emph{local} translation in spacetime, the conserved quantity for the present symmetry should be called the \emph{extended energy-momentum} tensor $\hat T_{A B}$, which is obtained from the variation of $\ell a_m+\ell a_\mathrm{id}=f_m \hat L$ with respect to the metric tensor variation, $\delta \hat g^{AB}$:
 \begin{equation}
 \hat{T}_{A B} \delta \hat g^{AB}=-2 \delta (\ell a_m+\ell a_\mathrm{id})+\hat{g}_{AB} (\ell a_m+\ell a_\mathrm{id}) \delta \hat g^{AB}.
 \end{equation} 
 
 Now, the \emph{Einstein tensor} for the extended phase space is obtained from the variation of $\ell a_\mathrm{f}$ with respcet to $\delta \hat g^{AB}$:
 \begin{equation}
 \hat G_{AB}=\hat{Ric}_{AB}-\hat{R} \, \hat{g}_{AB}/2,
 \end{equation}
 and the \emph{Einstein equation} can be written also for the extended phase space,
 \begin{equation}
 \label{7Einstein}
 \hat G_{AB}=8\pi \hat G\, \hat{T}_{A B}.
 \end{equation}
 It is worth noticing that, if confirmed, we have just obtained gravitation and electromagnetism from a geometrical perspective.  A similar equation holds in the \emph{Projective Unified Theories} proposed by Schmutzer \cite{schmutz} since '80, where the extended energy-momentum tensor is replaced by an \emph{energy projector} divided into a \emph{substrate energy-momentum tensor} and a \emph{scalaric energy-momentum tensor}. \\ 
  However, when extending the dimensionality from four to seven it is possible to take into account many possibilities. We will show that the abelian nature of the gauge theory comes suddenly from the choice of the  $\gamma \in  S^1$ gyro-phase as coordinate of the velocity space but, anyways, the gauge theory could become non abelian by choosing other variables with different groupal properties from the gyrophase. The possibility to definitely separate in the extended phase space what belongs to spacetime and what to velocity space must be reformulated. It seems that the space-time is simply defined as the domain of variation of the e.m. fields, in such a way that we need an e.m. field for defining space-time variables.  Such route needs some care and it cannot be taken just now. We prefer to show the minimal five dimensional extension of gravitation explicitly using the guiding center coordinates. Such extension is sufficient to include electromagnetism. Moreover, the present description is facilitated by the work of KK, because most of the general relativity equations that we will soon encounter, have already been studied \cite{historyKK}.
\subsection{The minimal five-dimensional theory}

Instead of deriving the metric tensor from a variational approach, it is possible to settle the metric tensor directly. This can be less elegant but easier to do mostly because it has already been done. The original KK mechanism needs an extension of the dimensionality of space-time by only one dimension. Only five dimensions occur to display electromagnetism and gravitation. However, we have formulated an extension to seven, not five, dimensions of general relativity. This is too general for the present scope, but we have seen that in the single particle one-form (\ref{tildeLform}) there is only the variation of five coordinates: $z^a=(t,X,\gamma)$, with a world line coordinate $\tilde s= \tilde s(z^a)$, for $a=0,1,2,3,4$. In this subsection we re-formulate the lagrangian density, (\ref{PCEH}), in five dimensions and, after adopting the KK metric tensor, we prove the equation (\ref{laHE}), which is the last equation needed to get the wanted lagrangian density (\ref{Ldens}). \\
The KK mechanism is used  following the review articles \cite{kk1} and \cite{kk2}. Many books can be consulted for the computation of the Ricci tensor and \emph{Christoffel symbols}, but a particularly interesting note inherited with the KK mechanism is \cite{KKforkids}. If two (canonical) constants of motion coordinates are taken into account, then the description of the dynamic of a particle in the extended phase space can be \emph{reduced} from seven to five dimensions. For the guiding center description of motion such coordinates are the  energy, $\varepsilon$, and the magnetic moment, $\mu$, and we can divide the extended phase space in slices of reduced phase space with assigned $\varepsilon$ and $\mu$. This is allowed because the co-momenta are $w_A=(\mathcal{E}, -P,(m/e)\mu,0,0)$, where the zeros are just indicating the use of canonical coordinates in $z^A=(t,X,\gamma,\varepsilon,\mu)$. The one-form (\ref{PCform0}) is the same of (\ref{KKform}) which lives in five dimensions. We have indicated with the \emph{hat} a seven dimensional quantity, \emph{e.g} $\hat L(z^A, z^{\prime B})$, whilst with a \emph{tilde} a five dimensional one, \emph{e.g.} $\tilde L(z^a, z^{\prime b})$. The lagrangian in (\ref{tildeLform}), $\tilde L=P\cdot U-\mathcal{E} U^0-(m/e)\mu \gamma^\prime $, is always the same but it is now written with the metric tensor $\tilde g_{a b}$:
\begin{equation}
\tilde L=-\tilde g_{a b} w^a z^{\prime b},\qquad \mbox{for a,b=0,1,2,3,4}.
\end{equation}
Also the \emph{lagrangian distribution}, (\ref{PCEH}) can be considered into five dimensions:
\begin{equation}
\label{ella}
\ell a= f_m \tilde L - \frac{\tilde R}{16 \pi  \tilde G},
\end{equation}
being $\tilde R$ the five dimensional scalar curvature, and $\tilde G$ the five dimensional gravitational constant. In practice, $\tilde R/ \tilde G=\hat R/ \hat G$, as if we are considering \emph{flat} the space described by the canonical coordinates $\varepsilon$ and $\mu$.
It is worth noticing that although in five dimensions, all the quantities can depend also on $\varepsilon$ and $\mu$, \emph{e.g} the distribution function $f_m$ is always the distribution of masses in the whole extended phase space and it surely depends on $\varepsilon$ and/or $\mu$ if it describes an equilibrium \cite{me2}.
Even if the action is the same,  now $\sqrt{|\hat{g}|}$ should be decomposed into $\sqrt{|\hat{g}|}=\sqrt{|\tilde{g}|} \tilde{J}_\mathcal{P}$, where $\sqrt{|\tilde{g}|}$ is the square root of the absolute value of the determinant of the metric tensor $\tilde g_{ab}$, and $\tilde{J}_\mathcal{P}$ is the jacobian, not specified here, for measuring the density of states for assigned $\varepsilon$ and $\mu$. From (\ref{7action}) and (\ref{Velem}), in guiding center coordinates, the action is
\begin{equation}
S=\int \ell a \, \sqrt{|\tilde{g}|} \tilde{J}_\mathcal{P} \,dt \,d^3 X \,d\gamma \,d \varepsilon \,d \mu.
\end{equation} 
Finally, we use the following KK metric tensor:
\begin{equation}
\label{KKmetric0}
\tilde{g}_{ab} = \left| \begin{array}{cc}
g_{\alpha \beta}+\kappa^2 \varphi^2  A_\alpha A_\beta  & \kappa \varphi^2 A_\alpha   \\  \kappa \varphi^2 A_\beta  & \varphi^2 \end{array} \right|.
\end{equation}
\subsubsection{The action for the fields}
If the chosen signature is $\eta_{\alpha \beta}=\mathrm{diag}(+1,-1,-1,-1)$ then $\varphi^2< 0$. Differently from KK, setting $\kappa^2 \varphi^2=-k_G^2$ and $\varphi^2=   -k_G^2 (m/e)^4 \mu^2$ (or $\kappa\mu=(e/m)^2$), the metric tensor becomes:
\begin{equation}
\label{KKmetric}
\tilde{g}_{ab} = \left| \begin{array}{cc}
g_{\alpha \beta}-k_G^2  A_\alpha A_\beta  & -k_G^2 (m/e)^2 \mu A_\alpha   \\  -k_G^2 (m/e)^2 \mu A_\beta  & -k_G^2 (m/e)^4 \mu^2 \end{array} \right|.
\end{equation}
being $k_G$ a constant that will be specified below. From (\ref{g^ab}), it is possible to obtain the contravariant momenta.
\begin{equation}
w^a=\tilde g^{ab} w_b=   \left|  \begin{array}{c} U^\alpha \\ (e/m)(1/\mu)[1+(e/m)^2(1/k_G)^2+\tilde L] \end{array} \right|,  
\end{equation}
in such a way that, from $\tilde L=-w^a U_a$, it is obtained the constancy of $\tilde L$ in terms of important physical constants:
\begin{equation}
\tilde L=-1-\frac{1}{2k_G^2(m/e)^2}.  
\end{equation}
The latter metric tensor is used to compute the five dimensional scalar Ricci tensor (\ref{App}): $\tilde R=R-\kappa^2\varphi^2 F_{\alpha \beta}F^{\alpha \beta}/4$. Now, the field action is 
\begin{equation}
S_\mathrm{f}=-\frac{1}{16 \pi \tilde {G}} \int dt d^3X \sqrt{|\tilde{g}|} \left(R +\frac{k_G^2}{4} F_{\alpha \beta}F^{\alpha \beta} \right)\tilde{J}_\mathcal{P}d\gamma d\varepsilon d\mu,
\end{equation}
where $\sqrt{|\tilde{g}|}$ is $\sqrt{|\tilde{g}|}=\sqrt{-g} (m/e)^2 k_G \mu$. For obtaining the standard gravitational plus e.m. fields action, $k_G$ must be $k_G^2= 16 \pi G$, so that 
\begin{equation}
\tilde G= G \int (m/e)^2 k_G \mu \tilde{J}_\mathcal{P} d\gamma d\varepsilon d\mu.
\end{equation}
  and 
\begin{equation}
S_\mathrm{f}=-\int  \sqrt{-g} dt d^3X \, \frac{ R}{ 16 \pi G} - \int  \sqrt{-g} dt d^3X \, \frac{F_{\alpha \beta}F^{\alpha \beta}}{4}.
\end{equation}
It is worth to note that the single particle interaction density lagrangian comes to be:
\begin{equation}
\tilde L_\mathrm{ime}=-\frac{e^2}{32\pi m^2 G}=-\frac{\alpha_\mathrm{fs}}{32\pi}\frac{\lambda^2_c}{\ell_p^2}.  
\end{equation}
where $\alpha_\mathrm{fs}=e^2/\hbar$, $\lambda_c=\hbar/m$ and $\ell_p=\sqrt{\hbar G}$.

In this way, we have obtained the lagrangian density in (\ref{Ldens}) from the five dimensional lagrangian (\ref{ella}). It is worth noticing that, even if the terms in the lagrangian density (\ref{Ldens}) are the desired ones, they are referring to fields on $(t,X)$ where $X$ is the guiding center position and it doesn't indicate the position of a particle. This is an effect of the misleading symmetry. The problem is that once we have integrated the lagrangian density, expressed in guiding center coordinates, on the velocity space, we have lost the possibility to know where the particles effectively are. This means that the present theory is \emph{non local}. Fortunately, such non-locality helps for the consistency of electrodynamics, \emph{e.g.} the problem of \emph{self-energy}, or \emph{self-interaction}, is promptly solved once a non-locality property is assumed. Moreover, we already know that, at some scale, an indetermination principle should be invoked. The relation between the misleading symmetry and the quantum non-locality property should be investigated. A simple guess is the following. In the \emph{Bhomian} formulation \cite{Bohmbook} of quantum mechanics the two ingredients are: \emph{strange} trajectories and non locality. We can easily prove that strange trajectories can be obtained from ad hoc e.m. field and that the property of non-locality has been just been obtained. However, a more precise draft on the relation between the present derivation and quantum mechanics is described in the next section.  \\
\subsubsection{Comments on the novel KK mechanism}
 The KK mechanism was discarded as a possible true mechanism of Nature because it holds many problems. The standard doubts refer to the reason for applying the cylinder condition, which is at the origin for explaining compactification. 
 Another problem is the compactified scale length of the order of the Planck length, $\ell_p$. Such scale length is inconsistent with the observed masses of elementary particles.
  Other approaches without these two ingredients, compactification and cylinder condition, are commonly less considered. However,  all the problems are inherited to explain why the fifth dimension is unobservable \cite{kk1,kk2}. In the present case, this is not a problem, because the fifth dimension is \emph{measurable}, being a physical meaningful and observable (not compactified) variable. The KK mechanism can be extended to include more species, more than five dimensions, generalized to include the \emph{cosmological constant} (see appendix A) and,  most importantly,  it is known to satisfy the \emph{Weyl} transformation \cite{kk1,kk2}. For simplicity, we don't examine these interesting extensions of the theory. Moreover, the present approach, that starts from the \emph{Lorentz' force law}, is completely \emph{Newtonian}. Thus, another big problem, as it happens with general relativity, will be its re-formulation  within the quantum mechanical rules. The procedure for obtaining a quantum mechanical description is even more difficult because we have explicitly used some issues that are not allowed in quantum mechanics, as the non-canonical hamiltonian description of motion and a gauge function which is not only defined over the space-time. Without an extension of the present theory to quantum mechanics it is not possible to accept the present theory. Similarly, for example, without a quantum reasoning it is not possible to deduce the scale of masses of the elementary particles. With respect to the latter remark an intriguing coincidence clearly  appears. If an indetermination principle is applied the fact that we have considered the $5^{th}$ dimension belonging to the velocity space should set the length scale of the extra dimension equal to the Compton length, not to the Planck length. Immediately, it is recognized that with the present, now compactified, KK mechanism,  also the scale of masses assumes the correct value. With an indetermination principle, the extra-dimension scale length becomes important because it is not possible anymore to know, at the same time, both the position and the velocity of the particle. Within quantum mechanics, it becomes forbidden to observe a gyro-radius below of the order of the Compton length. \\
  The extension of the present theory to quantum mechanics will be described in another work which is in preparation. However, here it can be roughly shown how the present approach is not  too much in conflict with quantum mechanics,  thanks to the misleading symmetry and the instability of the guiding centers due to electromagnetic fluctuations. 
 \section{Stochastic gyrocenter transformation}
 In this section we introduce quantum rules without following the orthodox way. The probabilistic concepts that pertain to the quantum world are shown to be consequences of e.m. fluctuations. 
 Several studies concerning the relation of quantum mechanics and stochastic processes are described in books like \cite{feynmann,Bohmbook,nagasawa,deLapena}. Others suggested lectures with many correspondences to the present derivation are in \cite{varma} and \cite{derak}. However, the present description is novel because it is applied to the gyrocenter, instead of considering the particle motion, when the gyrating particle solution is considered in the presence of e.m. stochastic fluctuations, that has never been studied. 
 \subsection{The stochastic gyro-center} 
The e.m. fluctuations are commonly considered separately from the guiding center description. There is a simple reason for this, indeed in non relativistic regime, it is possible to separately take into account the time behavior of the e.m. fields. In plasma physics applied to laboratory plasmas this is almost the case, because the guiding center approach is used for describing particles motion in the background \emph{equilibrium} e.m. fields $(E_0,B_0)$ that don't explicitly depend on time. The effective e.m. fields will be affected by changes induced by a redistributions of charges in the plasma. Such changes are e.m. fluctuations and they are particularly difficult to model because they are caused by collectives phenomena. However, in plasma physics modeling, the current approach \cite{brizard} is to give a spectral behavior to such fluctuations in such a way that after various efforts a dispersion relation is obtained.  The dispersion relation is known as the \emph{generalized fishbone-like} dispersion relation \cite{zonca}. Differently, here we consider stochastic fluctuations of the e.m. field. Moreover, given the stochastic nature of the e.m. fluctuations, we are inducted to separately consider the guiding center from the gyro-center description.
The single charged particle non relativistic Lagrangian is the same, but let us distinguish what is deterministic from what is stochastic:
\begin{eqnarray}
\nonumber
&&L=\dot {x}^2/2+(e/m)A_0(t,x)\cdot \dot{x}-(e/m)\Phi_0(t,x)+\\
&&+(e/m)\delta_\xi A\cdot \dot{x}-(e/m)\delta_\xi \Phi,
\end{eqnarray}
where $A_0^\alpha=(\Phi_0,A_0)$ is the (deterministic) \emph{four}-vector potential and $\delta_\xi A^\alpha=(\delta_\xi \Phi, \delta_\xi A)$ is the stochastic \emph{four}-vector potential fluctuations. The e.m. fluctuating fields could be written as $\delta_\xi E=-\partial_t \delta_\xi A -\nabla \delta_\xi \Phi$ and $\delta_\xi B=\nabla \times \delta_\xi A$, but some care should be considered when applying the stochastic differential calculus on such stochastic quantities.\\
The gyrocenter description is easily obtained from the guiding center description, because the effects of fluctuations will be easily reflected on the guiding center transformations that become
\begin{equation}
\bar X= X+\xi,
\end{equation}
where $\bar X$ is the gyro-center position, $X$ is the guiding center position and $\xi$ is the guiding center displacement.  Similarly for the velocity, 
\begin{equation}
\bar V= V+\delta_\xi V,
\end{equation}
where $\bar V$ is the gyro-center velocity, $V$ is the guiding center velocity and $\delta_\xi V=\dot \xi$ is the guiding center velocity displacement.
 Concerning stochastic processes, it is better to write the latter equation with the increments instead of the derivatives:
\begin{equation}
d\bar X=V dt+ d \xi.
\end{equation}
Moreover, being $V=v-\sigma$, the gyro-center increment $d \bar X$ is rewritten as
\begin{equation}
\label{forw}
d\bar X=(v-\sigma)dt+ d \xi,
\end{equation}
where the guiding center velocity, $V$, is the mean gyro-center velocity, and $d \xi$ at time $t$ is independent of $\bar{X}$ for a time before $t$.
In stochastic differential calculus  the limit $dt \to 0$ should be considered with care and it is meaningful to define two kinds of derivatives. The forward derivative:
\begin{equation}
D \bar X= \lim_{dt \to 0^+} \left \langle \frac{\bar X(t+dt)-\bar X(t)}{dt} \right \rangle,
\end{equation}
and the backward derivative
\begin{equation}
D_\star \bar X= \lim_{dt \to 0^-} \left \langle \frac{\bar X(t)-\bar X(t-dt)}{dt} \right \rangle,
\end{equation}
Here, $D \bar X=V$ and $D_\star \bar X=V_\star$.
In such a way that they are coincident, $V=V_\star$, when $\bar X$ is differentiable.
Thus, the stochastic process should be characterized also by the backward increments that can be written as
\begin{equation}
\label{back}
d\bar X=V_\star dt+ d \xi_\star,
\end{equation}
where $d \xi_\star$ at time $t$ is independent of $\bar{X}$ for a time \emph{after} $t$.
In general, $V_\star=v-\sigma_\star$.
 Finally, the stochastic process $\xi$ is considered a simple \emph{Wiener} process with
\begin{equation}
\langle d\xi \rangle= \langle d\xi_\star \rangle=0,
\end{equation}
and
\begin{equation}
\label{diff}
\langle d\xi_i d\xi_{\star j} \rangle=2\nu_p\delta_{ij}dt,
\end{equation}
with $d\xi_i$ and $ d\xi_{\star j}$ specifying the cartesian components of the stochastic vectors $d \xi$ and $d\xi_{\star}$, respectively.  
The constant $\nu_p$ indicates the product of a length times a velocity and coincides with the diffusion coefficient of the stochastic process. It is worth noticing that the origin of such diffusive process is due to the e.m. fluctuations. The implicit reason for such fluctuations are the absorbed and/or emitted radiation by the charge, its motion becomes \emph{markovian}, as for the \emph{brownian} particle. If we can ruled out the radiation, then the behavior could be different, for instance $\nu_p=0$. However, here we will always consider the presence of an e.m. field and, at least, one charge. The dynamics of a charge cannot correctly be described if separated from the e.m. field, that implies $\nu_p\neq 0$.  \\
It is possible to associate two Fokker-Planck (FP) equations to the stochastic process. For a probability density function, $f$, the forward FP equation is:
\begin{equation}
\partial_t f=-\nabla \cdot (Vf) +\nu_p \Delta f,
\end{equation}
or
\begin{equation}
\label{fp_for}
\partial_t f+V\cdot \nabla f=-f \nabla \cdot V +\nu_p \Delta f.
\end{equation}
Similarly, the Fokker-Planck equation for the backward process is 
\begin{equation}
\partial_t f=-\nabla \cdot (V_\star f) -\nu_p \Delta f,
\end{equation}
or 
\begin{equation}
\label{fp_back}
\partial_t f+V_\star\cdot \nabla f=-f \nabla \cdot V_\star +\nu_p \Delta f.
\end{equation}
From the sum and the difference of the two Fokker-Planck equations,
\begin{equation}
\label{conteq1}
\partial_t f =-\nabla \cdot \left(f\frac{V+V_\star}{2}\right).
\end{equation}
From subtracting the two FP equations,
\begin{equation}
\nabla \cdot \left( f\frac{\sigma-\sigma_\star}{2}\right)  +\nu_p \Delta f=0.
\end{equation}
From the latter, \emph{Nelson} argued the following particular solution for $\sigma-\sigma_\star=2 u_\mathrm{N}$, where the \emph{osmotic velocity} is
\begin{equation}
u_\mathrm{N}=-\nu_p \nabla \log f,
\end{equation}
as in \cite{nelson} apart from the minus sign.  Moreover, it is possible to define the Nelson's \emph{current velocity}:
\begin{equation}
v_N=V+u_N.
\end{equation}
The current velocity is the gyro-center velocity when $\dot \xi$ is replaced by $u_\mathrm{N}$. Obviously, when fluctuations are neglected, the gyro-center velocity becomes the guiding center velocity. \\
If the backward process is realized with $V_\star=v_\mathrm{N}+u_\mathrm{N}$, then the \emph{continuity equation}, from  eq.(\ref{conteq1}), is 
\begin{equation}
\label{conteqN}
\partial_t f =-\nabla \cdot (fv_\mathrm{N}),
\end{equation}
 This is the reason for appropriately calling the gyro-center velocity, $v_\mathrm{N}$, as the \emph{current velocity}. 
\subsubsection{the straight and uniform magnetic field with fluctuations}
In a straight and uniform magnetic field, we have seen that $\sigma=v_\perp$, which can be opportunely written as $\sigma=-\rho_L\omega_c(e_1 \sin \gamma+e_2 \cos \gamma)=\rho^2_L\omega_c \nabla \gamma$, if no fluctuations are considered. Here, $\gamma$ is always the gyro-phase, with  $\nabla \gamma=e_\gamma/\rho_L$, and $\omega_c=(e/m)|B|$. For such case without fluctuations, $\nabla f=0$, or $f=1$ , ensuring determinism. Thus, by introducing fluctuations that modify the \emph{effective} velocity of the charge, $v$, and the velocity $\sigma$. In such way that $v= v_N+\sigma-u_N=V+\sigma$.  It is worth noticing that he velocity  $v$ is an effective velocity, which is very useful because both the true velocity and the true e.m. field acting on the charge are unknown. Fluctuations add a term, the osmotic velocity $u_\mathrm{N}$, to $\sigma$ that becomes:
\begin{equation}
 \sigma=-\nu_p \nabla \log f+\rho^2\omega_c \nabla \gamma,
\end{equation}
where the Larmor radius is not anymore constant and it has been substituted with the gyroradius $\rho_L\rightarrow |\rho|$. Thus, $v=v_N+\rho^2\omega_c \nabla \gamma$. The backward velocity, $- \sigma_\star$ is obtained changing the direction of $\gamma$ (or the sign of the charge): $- \sigma_\star= -\nu_p \nabla \log f-\rho^2\omega_c \nabla \gamma$. In such a way that $ \sigma+ \sigma_\star=2 \rho^2\omega_c \nabla \gamma$. 
Always in the straight and uniform equilibrium magnetic field case, by taking $f \propto 1/(\pi \rho^2)$, that means that the probability for finding the particle is, roughly, inversely proportional to the area of the disc of radius $|\rho|$ \footnote{Another, more general, choice is $f^{-1} \propto \pi^n \rho^{2n}/\Gamma(n/2+1)$, which is the volume of a $n$-ball, with $n \in \mathbb N$.}. Now,  the divergency of $\nabla \cdot[f(\sigma+\sigma_\star)]=0$, being $\gamma$ an angle so that $\nabla^2\gamma=0$. In such case the continuity equation reads 
\begin{equation}
\label{conteq}
\partial_t f =-\nabla \cdot (fv),
\end{equation}
being $\nabla \cdot[f(V+V_\star)]=2\nabla \cdot (fv)-\nabla \cdot[f(\sigma+\sigma_\star)]=2\nabla \cdot (fv).$
 The \emph{gyro-velocity},
 \begin{equation}
\sigma=\frac{2\nu_p}{|\rho|} e_\rho+|\rho|\omega_c e_\gamma,
\end{equation} 
 is maintained perpendicular to the equilibrium magnetic field. The product $\rho \cdot \sigma=2\nu_p$ is constant, so that if  the radius of the disc, $|\rho|$, increases (\emph{e.g.} absoption of radiation) then the radial velocity decreases.  Thanks to the gyrating part, the overall  velocity, $|\sigma|$, increases. On the contrary, if $|\rho|$ decreases (\emph{e.g.} emission of radiation), the gyrating part of $\sigma$ becomes negligible with respect to the radial velocity that explodes as $\sim 1/|\rho|$. Such remarks, even if obtained in a non relativistic treatment, allows to roughly deduce the order of magnitude of the constant $\nu_p$ if a minimum value of $|\rho|$ and, correspondingly, a maximum velocity is conceived. Let's indicate the minimum radius, corresponding to the diffusion length, with $\lambda_c$, \emph{then}  a maximum velocity is obtained and indicated with $c \approx 2\nu_p/\lambda_c$. \\
The gyro-phase symmetry of the system is maintained so that the magnetic momentum is conserved, from equation (\ref{omegac}):
\begin{equation}
\mu=\frac{(e/m)\sigma^2}{2\omega_c}=\frac{(e/m)\omega_c \rho^2}{2}+\frac{2(e/m)\nu_p^2}{\rho^2\omega_c},
\end{equation}
where the contribution of fluctuations with respect to the standard magnetic momentum is evident from the appearance of the factor $\nu_p$ in the second term on the right hand side. As for the gyrating velocity, also the magnetic moment is never vanishing in the presence of fluctuations. Finally, the energy per unit mass is
\begin{equation}
\varepsilon=\frac{v^2_\|}{2}+\frac{\omega^2_c \rho^2}{2}+\frac{2\nu_p^2}{\rho^2}
\end{equation} 
 with a \emph{zero field point} energy per unit mass \cite{boyer} written as $\varepsilon_\mathrm{zfp}=2(m/e)\mu_B \omega_c$, and estimated to be, if $v_\|=0$ and ($|\rho|=\lambda_c, \nu_p \approx c\lambda_c /2)$: 
 \begin{equation}
 \varepsilon_\mathrm{zpf}\approx \left(\frac{\lambda^2_c \omega_c}{2}+\frac{c^2}{2\omega_c}\right)\omega_c.
\end{equation}
Thus, $\mu_B=(e/m)(\lambda^2_c \omega^2_c+c^2)/(4\omega_c)$.
Finally, the prestige is the following, if $\lambda^2_c \omega^2_c=c^2$ then the minimum energy of the charge (times the mass) is  $\mathrm{E}_\mathrm{zfp}\approx m c^2$. The surprise is that instead of being the energy of the particle at rest, in the present case, it is the guiding center which is at rest. Above all, the energy $m c^2$ has been obtained without a relativistic approach, but with a magnetic field $|B|=(m/|e|)(c/\lambda_c)$. This is not the only surprise, indeed,  if $\nu_p=\hbar/(2m)$ (as in \cite{nelson}) then $\lambda_c=\hbar/(mc)$ is the \emph{Compton} length (which means $|B|=m^2c^2/(|e|\hbar)$).   Now,
\begin{equation}
\rho \cdot (m\sigma) = \hbar,
\end{equation}
If you introduce $\Delta x=|\Delta x| e_\rho$, with $|\Delta x| \ge |\rho|$ and $\Delta p=|\Delta p|\sigma/|\sigma|$ with $|\Delta p|\ge |\sigma|$, as representative estimators of the indetermination of the position and of the velocity, respectively, of the charge with respect to the gyro-center, then
\begin{equation}
\Delta x \cdot \Delta p \ge \hbar 
\end{equation}
which is similar to the \emph{Heisenberg} indetermination principle. The explanation, with respect to the \emph{Copenhagen} interpretation, is quite different. The charge is moving with a newtonian deterministic motion, the gyro-center is moving with a stochastic motion and, due to such stochasticity, it is not allowed to know exactly the position and the velocity of the gyrocenter with respect to the charge and/or viceversa. Within the limit imposed by the indetermination principle the gyrocenter and the charge are undistinguishable entities.  In the following, the appellation of \emph{elementary particle} will be shown to be better suited for the stochastic gyro-center than for the charge. \\
 The zero field point energy (times the mass) becomes
\begin{equation}
\mathrm{E}_\mathrm{zfp} \approx \hbar \omega_c,
\end{equation}
about twice w.r.t. the one obtained from a quantum oscillator, but what is exactly needed for obtaining the \emph{black-body spectrum} from the \emph{Planck} distribution. Moreover, with the latter z.f.p. energy, the energy of the charge, $E_2=m\varepsilon$, with respect to the energy of the guiding center, $E_1=mv_\|^2/2$, is: $E_2-E_1=\hbar \omega_c$. The former is similar to the \emph{Bohr's frequency relation}, but it is a consequence of equation (\ref{fotoel}) when $(m/|e|)\mu=\hbar/m$.
It is worth noticing that such results have been obtained without introducing quantum mechanics or special relativity issues.\\
It is worth noticing that the non vanishing magnetic moment, due to the e.m. fluctuation is estimated to  $\mu_B=(|e|/m^2)\hbar/2$ which is the \emph{Bohr magneton}.
Moreover, asking for $f$ to be inversely proportional to the disc with radius equal to the radial position of the charge, we are constructing a measure for determining the probability of finding the unknown position of the particle. Is it possible that such construction leave us close to the \emph{Born} interpretation? 
\subsubsection{the closed magnetic field line with fluctuations}
An interesting behavior is seen if the canonical magnetic field line is closed. In the present paragraph we analyze the behavior of a charge when the canonical magnetic field is closed into a circle with radius $a_B=\lambda_c/\alpha_\mathrm{fs} \gg \lambda_c$. In the next paragraph, we will consider a system of an electron and an ion when the canonical magnetic field is closed into an \emph{invariant tori}. It is worth to note that it is important to have, at least, $\alpha_\mathrm{fs}^{-1}=O(10^2)$ for closing the circle without changing too much the former results with the straight magnetic field case. In fact, for a charge moving on a circle with a small radius, $\lambda_c$, the magnetic field is still sufficiently straight and uniform even if the field line is closed in a circle of radius $O(10^2)$ bigger than $\lambda_c$. A charge moving circularly with a radius $\lambda_c$ and velocity $\lambda_c \omega_c$, gives rise to a magnetic field which is almost straight in the vicinity of the charge but that it closes in a circle when the effects of the charge are mostly reduced. If $\omega_c=c/\lambda_c$ then the radius of the magnetic field line passing close to the center of the circle, and for which the motion of the charge is mostly the same as if it would straight, is $\lambda_c/\alpha_\mathrm{fs}$.\\
Thus, let's take $v_\|\rightarrow V_b=a_B \dot \theta$, where $\theta$ is a poloidal angle. Now, the charge is moving with an orbit on the surface of a torus of radius $a_B$  and described by the two angles, $\gamma$ and  $\theta$. The orbit is closed \emph{e.g.} if, given $n \in \mathbb N$ then  $n \dot \theta= \omega_c$. The closure can also happen on other tori with a radius greater then $a_B$;  such tori where the orbit are closed loops  are called \emph{invariant tori}. It is worth noticing the appearance of a diophantine relation.
Moreover, if  both the velocity of the charge in the \emph{Larmor} circle and of the guiding center in the poloidal circle is $c$, then $\alpha_\mathrm{fs}=1/n$ is the inverse of an integer. 
 The former picture is an approximation because the effects occurring when the cylinder is closed onto a tori has not been properly taken into account.   
\subsubsection{A toroidal magnetic configuration}
A picture very close to the \emph{de Broglie}'s model but with a reasoning applied to an \emph{invariant tori} on a 3D space, instead  to a simple closed string on a 2D surface, is suggested, here, once the system made by an \emph{electron} with mass $m_e$, and an \emph{ion} with mass $m_A$ and charge $Z|e|$ is considered. Such analysis is done for considering a rough relation between the toroidal magnetic configuration, as seen in tokamaks, and some aspects of the \emph{Bohr} atomic model. Only in the following section, a rigorous non relativistic atomic model can be addressed by deriving \emph{Schr\"odinger} equation. However, a suggestive idea explaining some old  disputes on the first appearances of quantum behaviors, is shown (without quantum mechanics). In the presence of an axisymmetric  magnetic field, described by (\ref{magnf}), the guiding center of the electron is given by equation (\ref{VaxiB}):
\begin{equation}
V_e=\lambda \frac{e}{m_e}\nabla \mathcal{P}_\phi \times \nabla \phi + \frac{e}{m_e} (\mathcal{P}_\phi-\psi_p )\nabla \phi,
\end{equation}
being $\lambda=-\psi_p/F$. If $\mathcal{P}_\phi=\psi_p+\lambda\mathcal{F}$ then the former velocity can be rewritten as
 \begin{equation}
V_e=\frac{e}{m_e} \frac{\mathcal{P}_\phi-\psi_p}{\mathcal{F}}\left(\nabla \mathcal{P}_\phi \times \nabla \phi + \mathcal{F}\nabla \phi \right).
\end{equation}
In parenthesis the magnetic field $B_c= \nabla \mathcal{P}_\phi \times \nabla \phi + \mathcal{F}\nabla \phi$ is rewritten with the \emph{Clebsh} representation as:
\begin{equation}
 B_c=\nabla \mathcal{P}_\phi \times \nabla (\phi-q_\mathrm{sf}\theta),
\end{equation}
 where $\theta$ is the generalized poloidal angle and 
 \begin{equation}
 q_\mathrm{sf}=\frac{B_c\cdot \nabla \theta}{B_c\cdot \nabla \phi}
 \end{equation}
  is the \emph{safety factor}. Now, $\mathcal{P}_\phi$ is the poloidal magnetic flux of $B_c$. It is possible to introduce a flux radial coordinate, $r$, in such a way that $\mathcal{P}_\phi \propto r^2$, which means that we are considering nested poloidal surfaces with circular cross sections. We also consider the presence of a positive charge, $Z|e|$, which is moving toroidally with the same toroidal component of the electron guiding center velocity, 
For describing the effective velocity of the electron, as for the ion, we should add to the guiding center velocity also the gyro-velocity, $\sigma$, with the osmotic velocity. However, we only wish to consider a particular case that reminds the old but always fascinating  Bohr's atom model.  It is chosen a very strange (never seen in tokamaks) safety factor with the following dependency on $r$:
\begin{equation}
q_\mathrm{sf}=\sqrt{r/a_B}, \qquad \mbox{ with $q_\mathrm{sf}\ge1 $}.
\end{equation}
Thus,  when the canonical magnetic field lines are closed (the guiding center orbits are closed, too), the resonant magnetic flux surfaces, $\mathcal{P}_{\phi.\mathrm{res}}$, are determined by the condition $\sqrt{r/a_B}=n \in \mathbb N$, or:
\begin{equation}
r=n^2 a_B,
\end{equation}
in such way that $\mathcal{P}_{\phi.\mathrm{res}}\propto n^4$.\\
Thus, only for some diophantine values of $r$ the guiding centers are resonants. Here, the question is if the magnetic field that allows the guiding center of the electron to move on invariant tori can, or cannot, be generated by the same electron and ion that we are describing. It is not easy to answer but what it can be said is that if it is chosen the reference frame where the toroidal guiding center velocity of the electron is null, then we arrive at the simple description of an electron moving circularly around a positive ion. In such reference frame, the ion is fixed. Concerning the electron, its motion is due to the cylindrical  symmetry of the system and to the electric field generated by the central positive ion. The balance of the electric field with the centripetal motion is 
\begin{equation}
\frac{Ze^2}{r^2}=m_e r \dot \theta^2,
\end{equation}
which means
\begin{equation}
 \dot \theta=\sqrt \frac{Ze^2}{m_er^3}=\frac{1}{n^3} \sqrt \frac{Ze^2}{m_e a_B^3}.
\end{equation}
The angular momentum, $L_\phi=m_e r^2 \dot \theta$ is  proportional to the number of poloidal cycles, $n$, that are necessary to close the orbit in the tori:
\begin{equation}
L_\phi=n \sqrt {m_eZe^2 a_B}.
\end{equation}
Finally, if $a_B=\hbar^2/(m_eZe^2)$ is the \emph{Bohr}'s radius, then the angular momentum is quantized:
\begin{equation}
L_\phi=n\hbar,
\end{equation}
which is the \emph{Bohr-Sommerfield} rule.
\\
Even if the former examples are somehow suggestives, the analysis is too rough and inappropriate for the delicateness of the problem. In the next section we abandon those simple cases for addressing a correspondence between stochastic gyrokinetic and quantum mechanics.
%
\\
\subsection{Nelson quantum mechanics}
Finally, the acceleration of the gyro-center is considered as in \cite{nelson}:
\begin{equation}
a_N=\frac{D D_\star+D_\star D}{2} \bar X.
\end{equation}
There are other possibilities on defining an acceleration but in this work they are not taken into account. Once the derivative is applied to $D \bar X=V$ and $D_\star \bar X=V_\star$, we find the Nelson's  acceleration:
\begin{equation}
\label{Nelsona}
a_N=\partial_t v_N+v_N\cdot \nabla v_N-u_N\cdot \nabla u_N+\nu_p\nabla^2u_N,
\end{equation}
It is worth noticing that $\dot v_N \neq a_N$, if fluctuations are considered. In other words, the trajectory of a particle is different if $f\neq1$ and $\nu_p\neq 0$. The idea of Nelson was to associate such discrepancy, that depends on the presence of fluctuations, with the quantum mechanical formulation. \\
In our case $a_N=(e/m)(E+v_N\times B)$ and, with the same procedure described in section II, it is possible to arrive at a modified velocity law equation:
\begin{eqnarray}
\nonumber
&&\partial_t p_N+\nabla \varepsilon_N-v_N\times \nabla \times p_N=\\
&&=(\nu_p^2/2)\nabla (\nabla \log f)^2+\nu_p^2\nabla \nabla^2 \log f,
\end{eqnarray}
being $p_N=v_N+(e/m)A$ and $\varepsilon_N=v_N^2+(e/m)\Phi$. However, a simple transversal electric field, $E_t=-(m/e)\partial_t p_N-(m/e) \nabla \epsilon_p$, is obtained if 
\begin{equation}
\label{epsilonn}
\epsilon_p=v_N^2/2+(e/m)\Phi-(\nu_p^2/2) (\nabla \log f)^2-\nu_p^2 \nabla^2 \log f,
\end{equation}
then
\begin{equation}
E_t+v_N\times B_c=0,
\end{equation}
similarly to the Lorentz's force law case. It is worth noticing that the relation between $\epsilon_p$ and the Bohm quantum potential \cite{bohm}, 
\begin{equation}
\label{QBohm}
Q_B=-\nu_p^2 f^{-1} \nabla^2 f=\nu_p \nabla \cdot u_N-u_N^2,
\end{equation}
is $$\epsilon_p=\varepsilon_N+u_N^2/2+Q_B.$$ Now, by considering the gyrating particle solution with $B_c=0$, it means 
that $p_N$ is a gradient, which is written $p_N=\nabla S_N$ and $E_t=0$. For simplicity, let's take
\begin{equation}
 \epsilon_p=-\partial_t S_N.
 \end{equation}
Nelson has shown in \cite{nelson} that the continuity equation in (\ref{conteqN}) and the equation for the acceleration in (\ref{Nelsona}) gives the \emph{Schr\"odinger} equation once $a_N$ is substituted with the \emph{Newtonian} force per unit mass: $F=m a_N$, and $v_N=\nabla S_N-(e/m)A$. Nelson's approach suffers from the \emph{Wallstrom} criticism \cite{wallstrom} that we easily overcome defining the complex function 
\begin{equation}
\label{psi0}
\psi=\sqrt{f}e^{-i\gamma},
\end{equation}
where $\gamma=S_N/(2\nu_p)$ is always the gyrophase, and  being an angle it is multivalued as noticed by Wallstrom. It is worth noticing that the gyrating particle solution corresponds to the zitter-solution already described in section III (C.1). In fact, it has been recently noticed in \cite{derak} that the zittter-solution can overcome the Wallstrom criticism. 
From (\ref{psi0}) they are easily obtained the relations:
\begin{eqnarray}
f\partial_t \gamma&=& i\frac{\psi \partial_t \psi^\star-\psi^\star \partial_t \psi}{2},\\ 
f\nabla \gamma&=&  i\frac{\psi \nabla \psi^\star-\psi^\star \nabla \psi}{2},
\end{eqnarray}
and
\begin{eqnarray}
&&f\partial_t \log f=\partial_t f=\psi^\star \partial_t \psi +\psi \partial_t \psi^\star\\
&&f\nabla \log f=\nabla f=\psi^\star \nabla \psi +\psi \nabla \psi^\star.
\end{eqnarray}
Thus,
\begin{equation}
u_N=-\nu_p\frac{\psi^\star \nabla \psi +\psi \nabla \psi^\star}{\psi^\star \psi}
 \end{equation}
 and the stochastic gyrocenter velocity, $v_N=-2\nu_p\nabla \gamma-(e/m)A$, is
 \begin{equation}
 v_N=+i\nu_p\frac{\psi \nabla \psi^\star-\psi^\star \nabla \psi}{\psi^\star \psi}-(e/m)A.
  \end{equation}
  At the moment it doesn't occur to specify that the potentials are computed in $\bar X$, however it makes a certain difference. 
  \subsubsection{The Schr\"odinger equation from classical physics and stochasticity}
  Even if Nelson was clear in his derivation, we follow a different approach,  a constructive one, to reach the Schr\"odinger equation.
  It is here required that $v_N \cdot u_N=0$, which means that the stochastic gyrocenter velocity is perpendicular to the osmotic velocity due to fluctuations. Such choice is a requirement on $f$.
  From the guiding center velocity, $V=v_N-u_N$, it follows that $V^2=v_N^2+u_N^2=(v_N+i u_N)\cdot(v_N-iu_N)$.
  In terms of $\psi$ and $\psi^\star$, $V^2$ is below computed. Firstly
  \begin{equation}
  v_N-iu_N=\psi^\star \frac{-2i\nu_p \nabla-(e/m) A}{f} \psi,
  \end{equation}
  and
  \begin{equation}
  v_N+iu_N=\psi \frac{2i\nu_p \nabla-(e/m) A}{f} \psi^\star.
  \end{equation}
  Thus, the guiding center velocity squared is
  \begin{eqnarray*}
  &&V^2=f^{-1}[2i\nu_p \nabla-(e/m)A ] \psi^\star \cdot\\
  && \cdot [-2i\nu_p \nabla-(e/m)A] \psi=\\
  &&=2i\nu_p f^{-1}\nabla \cdot f(v_N+iu_N)+\\
  &&+\psi^\star f^{-1} [-2i\nu_p \nabla-(e/m)A]^2\psi.
    \end{eqnarray*}
    From the continuity equation (\ref{conteq}) and from the Bohm's quantum potential (\ref{QBohm}), $V^2/2$ is rewritten as
    \begin{equation}
    \frac{V^2}{2}=-i\nu_p f^{-1}\partial_t f-Q_B+\psi^\star f^{-1} \frac{[-2i\nu_p \nabla-(e/m)A]^2}{2}\psi
    \end{equation}
    Finally, it occurs only set all the pieces together, from (\ref{epsilonn}):
    \begin{eqnarray}
&&    \epsilon_p=V^2/2+(e/m)\Phi+Q_B=\\
\nonumber
&&= -i\nu_p f^{-1}\partial_t f +\psi^\star f^{-1} \frac{[-2i\nu_p \nabla-(e/m)A]^2}{2}\psi+(e/m)\Phi,
    \end{eqnarray}
    being $\Phi$ computed at the gyrocenter position, $\bar X$. Moreover, being $\epsilon_p=-2\nu_p\partial_t \gamma$, the following equation is easily obtained:
    \begin{eqnarray*}
  &&  -2\nu_p\partial_t \gamma+i\nu_p f^{-1}\partial_t f=\\
  &&=\psi^\star f^{-1} \frac{[-2i\nu_p \nabla-(e/m)A]^2}{2}\psi+(e/m)\Phi.
    \end{eqnarray*}
     If $\nu_p=\hbar/(2m)$, as already considered in the straight and uniform magnetic field case with fluctuations, then the Schr\"odinger equation is derived:
    \begin{equation}
    \label{schroe}
    2i\nu_p\partial_t \psi= \frac{[-2i\nu_p \nabla-(e/m)A]^2}{2}\psi+(e/m)\Phi \psi.
    \end{equation}
     Finally, if we set the minimum allowed magnetic moment to
    \begin{equation} 
    \mu=-(e/m)\nu_p=\mu_B,
    \end{equation}
     then the order of magnitude of the \emph{compactification} scale in the KK mechanism is the Compton length, $\lambda_c$, ensuring the correct mass scale for the elementary particles. Moreover, $\nu_p$ is the diffusion coefficient in (\ref{diff}), in such a way  that  the Wiener process is recognized to be universal (as already noticed by Nelson).
\section{Conclusions}

The non-perturbative guiding center transformation has been extended to the relativistic energies. Within the relativistic regime, the same equation (\ref{ohmslaw}) already seen in the non relativistic treatment \cite{me0}, is re-obtained. This has been called the \emph{velocity law}. Although the context is very different, the similarity with the \emph{ideal Ohm's law} has been shown and, some solutions of motion are studied in the light of the ideal Ohm's law. The covariant formalism has been adopted to better describe the relativistic behavior. For this reason a lagrangian approach is used for re-deriving the same equation  (\ref{ohmslaw}) in a covariant form.\\
Some important solutions of the velocity law are considered in section III. Here, the difference between the guiding particle solution in gyrokinetic-like ordering, in MHD-like ordering, and the gyrating particle solution, is shown.  All these solutions are practically identical to the non relativistic case, which have been analyzed in detail in \cite{me0}. The guiding particle solution is the one described by the \emph{fundamental equation} (\ref{claudio1}); the guiding center can be described by the same equation but having the magnetic moment different from zero. The guiding center reference frame has been finally defined in a geometrical sense as \emph{the reference frame where the particle moves in a closed orbit with a periodic motion}. The gyro-phase, $\gamma$, is the curvilinear coordinates along the closed loop trajectory and the magnetic moment is defined as the conjugate coordinate to $\gamma$. Thus,  the dynamics have been described in the guiding center coordinates, $z^A=(t,X,\gamma,\varepsilon,\mu)$, through the non-canonical hamiltonian mechanics developed by Cary and Littlejohn \cite{cary1983}. The \emph{Lagrange} and \emph{Poisson} tensors have been described for the non-perturbative guiding center transformation. The correspondence with the velocity law in seven dimensions is shown, (\ref{generalohm}). Moreover, a clear and known criteria to define when a dimensionality reduction is possible, is also reminded.\\

Furthermore, a general relativity approach for describing electromagnetism using the relativistic guiding center transformation is suggested. It is shown that the formalism of non-canonical hamiltonian mechanics is what is needed to extend the presence of electromagnetic dynamics to the general relativity formalism.  An \emph{Einstein's equation} (\ref{7Einstein}),  for the extended phase space can be settled for describing both the interactions: elelctromagnetism plus gravitation. Moreover, it has been proved that, for the guiding center coordinates, the relevant dynamics are five dimensional as for the original KK mechanism. The lagrangian density (\ref{Ldens}), which is used for describing both gravitation and electromagnetism, has been obtained. The metric tensor has been explicitly written in (\ref{KKmetric}). The gyro-phase coordinate, $\gamma$, is proposed to be the fifth KK coordinate. Thus, the extra-dimension is not an unobservable spacetime dimension but a \emph{measurable} coordinate of the velocity space used for  describing motion on the extended phase space. For this reason, the KK mechanism does't need of a compactification procedure, anymore.  If $\gamma \in S^1$, which is exactly obtained only for the \emph{non-perturbative} guiding center transformation, an \emph{abelian gauge theory} can be settled: electromagnetism is served on the gravitational banquet. The novelty of the present work relies on the fact that the geometry of the velocity space must be taken into account also for describing the same gravitational field acting on the particles. \\
In the last section, some speculative possibilities are taken into account. Once electromagnetic fluctuations are considered,  it is not allowed anymore to shrink  the gyroradius to zero. From the guiding center transformation to the stochastic gyrocenter one, it occurs to radically change the velocity law, which means that the gyrocenter moves differently with respect to the guiding center. Using Nelson's approach to quantum mechanics, answering to the Wallstrom's criticism and giving a physical justification to the fluctuations required by Nelson's approach, then it has been shown how to derive the Schr\"odinger equation (\ref{schroe}), with \emph{all} its implications.\\
 Finally, considering the diffusion coefficient as proportional to the \emph{Planck} constant and inversely proportional to the inertial mass, it has been shown that the scale of length for the extra dimension is the Compton length, instead of the Planck length. This is correct also from another kind of consideration. Thus, if the extra dimension belongs to the velocity space, the Heisenberg indetermination principle which forbids the contemporary knowledge of position and of velocity of the particle, led to a different scale length limitation, which is caused by the incommensurability between velocities and positions instead of by the unobservability of the 5th dimension. Thus, the length scale for the new \emph{compactification} scheme is fixed by the Compton length, ensuring the correspondence with the observed masses.\\
What emerges from such picture is that some quantum effects can be also explained, and not only interpreted, by the old classical mechanics.  Once the non perturbative guiding center and the stochastic gyrocenter transformations are applied to plasma physics then a field theory (on extended phase-space) approach can be, finally, well suited for solving nonlinearities. From the plasma lagrangian density in (\ref{somenew}) where "somethingnew" is substituted with the Hilbert-Einstein term on extended phase-space, the consequences and the differences with the standard formulation should be investigated. Once macro- and micro- behaviors will be described in a unified manner, then  the multi-scale non-linear problem encountered in tokamak physics can be reformulated with new tools; also for this reason, the non-perturbative guiding center and the stochastic gyrocenter transformations have been proposed. 
\begin{acknowledgments}
The author wishes to acknowledge S. Sportelli, S. Briguglio, E. Giovannozzi, F. Zonca and C. Cosentino for encouragement  and suggestions. This work has been carried out within the framework of the \emph{Nonlinear Energetic Particle Dynamics} (NLED) European Enabling Research Project, WP 15-ER-01/ENEA-03, within the framework of the EUROfusion Consortium and has received funding from the Euratom research and training programme 2014-2018 under grant agreement No 633053. The views and opinions expressed herein do not necessarily reflect those of the European Commission.
\end{acknowledgments}
\appendix
\section{Christoffel and Ricci in 5D}
\subsubsection{Christoffel symbols}
From
\begin{equation}
\tilde \Gamma^a_{bc}=\frac{1}{2}\tilde g^{ae} (\partial_c\tilde g_{eb}+\partial_b\tilde g_{ce}-\partial_e \tilde g_{bc}),
\end{equation}
with
\begin{equation}
\label{g^ab}
\tilde{g}^{ab} = \left| \begin{array}{cc}
g^{\alpha \beta}  & -\kappa  A^\alpha   \\  -\kappa  A^\beta  & \kappa^2 A^\alpha A_\alpha+1/ \varphi^2 \end{array} \right|,
\end{equation}
it is possible to compute all the components of the Christoffel symbol.
\begin{equation}
\tilde \Gamma^\alpha_{\beta \delta}=\Gamma^\alpha_{\beta \delta}+\frac{\kappa^2 \varphi^2}{2}g^{\alpha \eta} (A_\delta F_{\beta \eta}+ A_\beta F_{\delta \eta}), 
\end{equation}
\begin{eqnarray*}
&&\tilde \Gamma^4_{\beta \delta}=-\frac{\kappa}{2}A_\alpha \Gamma^\alpha_{\beta \delta}+\frac{\kappa^3 \varphi^2}{2} A_\delta A^\alpha F_{\alpha \beta}+\\
&&+ \frac{\kappa^3 \varphi^2}{2} A_\beta A^\alpha F_{\alpha \delta}+\frac{\kappa}{2}(\partial_\beta A_\delta+\partial_\delta A_\beta),
\end{eqnarray*}
\begin{equation}
\tilde \Gamma^\alpha_{\beta 4}=\frac{\kappa \varphi^2}{2}g^{\alpha \eta} F_{\beta \eta},
\end{equation}
and
\begin{equation}
\tilde \Gamma^\alpha_{4 4}=\tilde \Gamma^4_{4 4}=0.
\end{equation}
Last,
\begin{equation}
\tilde \Gamma^4_{\alpha 4}=\frac{\kappa^2\varphi^2} {2}A^\delta F_{\delta \alpha},
\end{equation}
and
 \begin{equation}
\tilde \Gamma^4_{\alpha 4}=\tilde \Gamma^4_{4 4}=0.
\end{equation}
\subsubsection{Scalar curvature in 5D}
From the components of $\Gamma$ the component $\tilde R_{44}$ of the Ricci tensor in 5D can be computed:
 \begin{equation}
 \tilde R_{4 4}=\tilde R^a_{4 a 4}=\tilde R^4_{4 4 4}+\tilde R^\beta_{4 \beta 4}=\tilde R^\beta_{4 \beta 4},
 \end{equation}
 with
 \begin{equation}
 \tilde R^\beta_{4 \beta 4}=\frac{\kappa^2\varphi^4}{4} F^{\alpha \beta}F_{\alpha \beta}-\frac{1}{2}\nabla^\alpha\nabla_\alpha \varphi^2
 \end{equation}
 The other components are
 \begin{equation}
 \tilde R_{\alpha 4}=\tilde R^a_{\alpha a 4}=\tilde R^4_{\alpha 4 4}+\tilde R^\beta_{\alpha \beta 4}=\tilde R^\beta_{\alpha \beta 4}
 \end{equation}
 with
 \begin{equation}
\tilde R^\beta_{\alpha \beta 4}=\frac{\kappa \varphi^2}{2} g^{\delta \eta} \nabla_\eta F_{\alpha \delta}+\kappa A_\alpha  \tilde R^\beta_{4 \beta 4}.
 \end{equation}
 Finally,
  \begin{equation}
 \tilde R_{\alpha \beta}=\tilde R^a_{\alpha a \beta}=\tilde R^4_{\alpha 4 \beta}+\tilde R^\delta_{\alpha \delta \beta}=\tilde R^\beta_{\alpha \beta 4},
 \end{equation}
 explicitly
  \begin{eqnarray}
  \nonumber
 &&\tilde R_{\alpha \beta}=R_{\alpha \beta}-\frac{\kappa^2\varphi^2}{2} g^{\delta \eta}F_{\alpha \delta}F_{\beta \eta}+\kappa^2A_\alpha A_\beta \tilde R_{44}+\\ 
 \nonumber
 &&+\kappa A_\alpha(\tilde R_{\beta 4}-\kappa A_\beta \tilde R_{4 4})+\kappa A_\beta(\tilde R_{\alpha 4}-\kappa A_\alpha \tilde R_{4 4})+\\
 &&-\frac{1}{2\varphi^2}\nabla_\alpha \nabla_\beta \varphi^2=\\
 \nonumber
 &&=R_{\alpha \beta}-\frac{\kappa^2\varphi^2}{2} g^{\delta \eta}F_{\alpha \delta}F_{\beta \eta}+\kappa^2A_\alpha A_\beta \tilde R_{44}+\\
 \nonumber
 &&+\frac{\kappa^2 \varphi^2}{2} g^{\delta \eta} A_\alpha \nabla_\eta F_{\beta \delta}+\frac{\kappa^2 \varphi^2}{2} g^{\delta \eta} A_\beta \nabla_\eta F_{\alpha \delta}
 \end{eqnarray}
 Thus, the scalar curvature in 5D is
 \begin{equation}
 \tilde R=R-\frac{\kappa^2\varphi^2}{4}F^{\alpha \beta} F_{\alpha \beta} -\frac{1}{\varphi^2}\nabla^\alpha \nabla_\alpha \varphi^2
 \end{equation}
 It is convenient to settle the following \emph{Klein-Gordon} equation with the \emph{Laplace-Beltrami} operator:
 \begin{equation}
 (\nabla^\alpha \nabla_\alpha +\Lambda) \varphi^2=0,
 \end{equation}
 being $\Lambda$ prop. to the cosmological constant. However, let's consider the simplest case with $\Lambda=0$. In such case,
 \begin{equation}
 \label{App}
 \tilde R=R-\frac{\kappa^2\varphi^2}{4}F^{\alpha \beta} F_{\alpha \beta}.
 \end{equation} 
\bibliography{DiTroiaPhysRevD17eurof}

\end{document}